\documentclass[preprint]{aastex63}

\usepackage{natbib}
\usepackage{amsmath}

\accepted{\today}
\submitjournal{ApJ}

\shorttitle{Habitable planetary orbits in stellar binary systems}
\shortauthors{Simonetti et al.}

\begin{document}

\title{Statistical properties of Habitable zones in stellar binary systems}

\correspondingauthor{Paolo Simonetti}
\email{paolo.simonetti@inaf.it}

\author[0000-0002-7744-5804]{Paolo Simonetti}
\affil{University of Trieste - Dep. of Physics, 
Via G. B. Tiepolo 11, 
34143 Trieste, Italy}
\affil{INAF - Trieste Astronomical Observatory,
Via G. B. Tiepolo 11,
34143 Trieste, Italy}

\author[0000-0001-7604-8332]{Giovanni Vladilo}
\affil{INAF - Trieste Astronomical Observatory,
Via G. B. Tiepolo 11,
34143 Trieste, Italy}

\author[0000-0002-7571-5217]{Laura Silva}
\affil{INAF - Trieste Astronomical Observatory,
Via G. B. Tiepolo 11, 
34143 Trieste, Italy}

\author{Alessandro Sozzetti}
\affil{INAF - Torino Astrophysical Observatory,
Via Osservatorio 20, 
10025 Pino Torinese, Italy}




\begin{abstract}
Observations of exoplanets and protoplanetary disks show that binary stellar systems can host planets in stable orbits. Given the high binary fraction among stars, the contribution of binary systems to Galactic habitability should be quantified. Therefore, we have designed a suite of Monte Carlo experiments aimed at generating large (up to $10^6$) samples of binary systems. For each system randomly extracted we calculate the intersection between the radiative habitable zones and the regions of dynamical stability using published empirical formulations that account for the dynamical and radiative parameters of both stars of the system. We also consider constraints on planetary formation in binary systems. We find that the habitability properties of circumstellar and circumbinary regions are quite different and complementary with respect to the binary system parameters. Circumbinary HZs are, generally, rare ($\simeq 4\%$) in the global population of binary systems, even if they are common for stellar separations $\lesssim 0.2$\,AU. Conversely, circumstellar HZs are frequent ($\ge 80\%$) in the global population, but are rare for stellar separations $\lesssim 1$\,AU. These results are robust against variations of poorly constrained binary systems parameters.
We derive ranges of stellar separations and stellar masses for which HZs in binary systems can be wider than the HZs around single stars; the widening can be particularly strong (up to one order of magnitude) for circumstellar regions around M-type secondary stars. The comparison of our statistical predictions with observational surveys shows the impact of selection effects on the habitability properties of detected exoplanets in binary systems.

\end{abstract}


\section{Introduction}
\label{sec:intro}

One of the aims of exoplanetary studies is to find habitable environments outside the Solar System. The habitability is most generally assessed by evaluating the climatological conditions that allow liquid water to be present on the  surface of rocky planets \citep{Kasting88}. Liquid water is essential for terrestrial life \citep[e.g.,][]{Westhof93} and, thanks to the hydrogen-bonding properties of the water molecule, provides an optimal molecular medium for any biochemistry sustained by genetic and catalytic molecules \citep{VladHassan18}. Despite the apparent simplicity of the liquid water criterion, quantitative assessments of habitability are challenging due to the difficulty of modeling the extreme climatological conditions that define the limits of planetary habitability, such as the runaway greenhouse instability or the maximum greenhouse \citep{Kasting93}. The problem is complicated by the fact that the climate and habitability of exoplanets are influenced by a broad range of stellar, orbital and planetary properties that are different from the terrestrial ones \citep{Silva17b,Ramirez19}. Among all possible climate factors, the insolation (i.e. the radiative  flux the planet receives from its host star) plays a prominent role in determining the surface planetary conditions \citep[see, e.g.][]{Vladilo15}. In conjunction with the spectral energy distribution (SED) of the central star, the insolation is commonly used to define the spatial extent of the circumstellar habitable zone (HZ) \citep{Kopparapu13a,Kopparapu14}.

Most studies of habitability are focused on planetary systems orbiting single stars, in particular Solar twins hosting an Earth-like planet. However, the advancements in the characterization of exoplanetary systems have shown that our system is just a single possible outcome among many different architectures \citep{Hatzes16}. The studies of planetary habitability must therefore take a broader perspective. In this respect, an important point to consider is that $\sim 50$\% stars in the field are in binary or even multiple star systems \citep[]{Duquennoy91, Kouwenhoven07, Bergfors10, Raghavan10, DucheneKraus13, Tokovinin14}. Therefore, the impact of stellar binarity must be taken into account in studies of planetary formation and habitability \citep[see, e.g.,][]{PilatLohinger19}.

The efficiency of planetary formation in binary and multiple systems with respect to single stars is debated, since the presence of multiple stars may affect the protoplanetary disk. This is true both for the circumstellar case \citep{Quintana07, JangCondel15, Thebault15, Rafikov15b} and for the circumbinary case \citep{Quintana06, Bromley15, Czekala19}. However, it is a fact that a relatively large number of planets in binary stellar systems have been discovered, both in circumstellar \citep[e.g.][]{Hatzes03,Barnes20} and in circumbinary \citep[e.g.][]{Doyle11,Kostov20} orbits. As of June 2020, based on the catalogue by \citet{Schwarz16}\footnote[4]{http://www.univie.ac.at/adg/schwarz/multiple.html}, 150 exoplanets have been found in 102 binary systems and 36 exoplanets in 26 multiple star systems. Currently, exoplanet-hosting stars in binary systems amount to about $\sim 2\%$ of the total, but considering the entire Galaxy this percentage could go up to $10-15$\% \citep{Raghavan06, Mugrauer09, Roell12}. The latter number provides the rationale for a systematic, statistical approach to assess the impact of stellar binarity on planetary habitability of the kind we address in this paper.

Only planets in stable orbits can benefit from climatological conditions that allow the long-term persistence of liquid water. Therefore a prerequisite of planetary habitability in binary systems is the existence of regions of dynamical stability  in the three-body system. The existence of dynamically stable regions is a simplification (for instance, stable planetary orbits could be forced to relatively high value of eccentricity), but it is nonetheless a good starting point. Numerical integrations of the general three-body problem indicate that regions of dynamical stability do exist in binary systems \citep{Harrington77, Rabl88, Dvorak88, Dvorak89, HW99, Doolin11, Quarles18, Quarles20}. Two main dynamical classes are possible: circumstellar orbits around one component of the system (S-type orbits) and circumbinary orbits around both stars (P-type orbits). For S-type orbits dynamical stability is achieved when the planetary semi-major axis is smaller than a critical value $a_\text{crit,S}$, whereas for P-type orbits the stability region is found outwards of a critical value $a_\text{crit,P}$. From  dynamical simulations of 10$^4$ binary orbits, \citet{HW99} extracted empirical expressions to estimate critical values of orbital stability as a function of the binary parameters. A reassessment of these calculations, based on a much larger number of simulations, covering 10$^6$ binary orbits and a broader space of dynamical configurations (e.g. non-coplanar planetary orbits), was recently performed by \citet{Quarles18,Quarles20}. The basic results concerning the location of the stability regions found by \citet{HW99} are confirmed by \citet{Quarles18,Quarles20}. It should be noted that there is not a complete distinction between stable and unstable regions: islands of stability can exist inside otherwise unstable regions and vice versa due to the interplay between competing orbital resonances. However, these resonant configurations would probably be the exception rather than the rule. The same is true for L-type orbits around the L$_4$ and L$_5$ Lagrangian points of the binary system \citep{Schwarz12} which are only relevant for a minority of systems \citep{HW99}. In the present work we focus on the habitability of coplanar S-type and P-type orbits and we adopt the analytical expressions for $a_\text{crit,P}$ and $a_\text{crit,S}$ provided by \citet{Quarles18} and \citet{Quarles20}, respectively.

The effects of insolation on the habitability of circumstellar and circumbinary regions in binary systems has been investigated by several authors \citep[]{Kane13, Kaltenegger13, Haghighipour13, Cukier19}. A common approach of these studies is to combine the flux from the two stars by properly weighting each contribution so to define radiative HZs for binary systems.
In some cases analytical expressions of the HZ incorporating the stability conditions have been derived \citep{Eggl12, Cuntz14, Wang17}. In general it is found that radiative HZs intersecting the stability regions can exist. The habitability around low-mass stars could even improve in binary systems with respect to the single star case, because tidal breaking between the two stars is expected to reduce the magnetic dynamo action and thereby the stellar activity that leads to erosion of the planetary atmosphere \citep{Mason13, Mason15, Zuluaga16}. Moreover, \citet{May16} and \citet{HaqqMisra19} demonstrated that atmospheric redistribution and surface heat capacity are able to contain surface temperature oscillations within less than few percent in a variety of cases. On the other hand, \citet{Popp17} found that the periodic radiative forcing due to the binary orbital mechanics have important effects on the planet's climate. In \citet{Yadavalli20} this forcing is studied in relation also to other binary parameters, most importantly the mass ratio between the components. \citep{Quarles19} found that the tilt of the planetary rotation axis is particularly complex to stabilize in binary systems a condition that can negatively affect the planetary habitability. Finally, circumstellar planets are subject to secular perturbations from the distant companion star which may undermine the dynamical stability of the radiative HZ in presence of a Jupiter-like planet external to the HZ \citep{Bazso19}.

Given the potential capability of  binary stellar systems to host habitable exoplanets, in the present work we introduce a statistical approach aimed at quantifying the circumstellar and circumbinary properties of habitability as a function of the stellar and orbital parameters of the systems. Starting from a set of statistical distributions constrained by observations of binary stars, we generate a large number of binary systems by means of Monte Carlo simulations. We limit our attention to systems with main-sequence stars, the most interesting from the point of view of long-term habitability. For each system, we calculate the location and extension of the circumstellar and circumbinary regions with long-term habitability conditions, both in terms of insolation and dynamical stability. The results, presented  in Section \ref{sec:results}, reveal important differences and peculiarities in the habitability properties of S-type and P-types regions. In Section \ref{sec:discussion} we discuss the impact of these results in planning observational searches for habitable planets in binary systems. In Section \ref{sec:conclusions} we draw the conclusions and discuss future extensions of this work.

\section{Methods} 
\label{sectMethods}

The main steps of our methodology can be summarized as follows. First we generate a Monte Carlo sample of binary systems.
For each  system of the sample we then calculate the boundaries of circumbinary and circumstellar  regions where (1) the planetary orbits are dynamically stable and (2) the insolation allows the long-term existence of surface liquid water. Finally we investigate the habitability of the circumstellar and circumbinary regions of the simulated sample as a function of binary system parameters. 

\subsection{Building the sample of binary systems}  
\label{sectMonteCarlo}

The sample of binary systems is generated by random sampling of probability density functions (PDFs) of stellar masses, orbital periods, mass ratios, and orbital eccentricities that we now describe in detail.

\subsubsection{The primary star mass function}
\label{sectPrimaryMass}
 
First, we draw the mass of the primary star ($m_\text{A}$) of the system using a mass function (MF), $\xi (\log m)$, which represents the number of stars per unit volume per logarithmic mass interval $d \log m$. To this end we use empirical relations derived for stars of the Galactic disk \citep{Chabrier03b}. For $m \leq 1$\,M$_\sun$ we adopt the log-normal distribution $\xi (\log m) = A \exp \{  - { (\log m - \log m_c)^2 / 2 \sigma_{\log m}^2}  \}$ with $A=0.086$ pc$^{-3}$ $(\log \mathrm{M}_\sun)^{-1}$, $m_c=0.079$\,M$_\sun$, and $\sigma_{\log m}=0.69$. For $m > 1$\,M$_\sun$ we adopt the classic power law suggested by \citet{Salpeter55}, $\xi (\log m) = A \, m^{-x}$. Since high-mass stars evolve significantly in the course of a Hubble time we use two different sets of parameters, one representative of the initial mass function (IMF) and the other of the present-day mass function (PDMF). In practice,  we adopt $A=4.43 \times 10^{-2}$ pc$^{-3}$ $(\log \mathrm{M}_\sun)^{-1}$ and $x=1.3$ for the IMF and $A=4.4 \times 10^{-2}$ pc$^{-3}$ $(\log \mathrm{M}_\sun)^{-1}$ and $x=4.37$ for the PDMF \citep{Chabrier03b}. The stellar mass is drawn in the range  $0.1 \leq m/(\mathrm{M}_\sun) \leq 1.5$, which is within the range of validity of the above distributions (the parametrization of the PDMF that we use is valid up to $m= 3.5$ M$_\sun$). Stars with $m >1.5 $ M$_\sun$ are not considered because they are less frequent and because they are less interesting for studies of habitability given the fact that their lifetime on the main sequence is small compared to the evolutionary timescales for terrestrial life as we know it.In Sec. \ref{sectOccurrence} we explore the impact of a possible spectral type-dependent occurrence of primary stars in binaries different from that of single stars.

\subsubsection{The distribution of stellar orbital periods}
\label{sectPeriods}

As a second step, we draw a value for the orbital period of the binary system in days, $P_b$, using the log-normal distribution suggested by \citet{Raghavan10}:
\begin{equation}
\xi(\log P_b)\propto\exp\{-{(\log P_b-\log\overline{P_b})^2/2\sigma_{\log P_b}^2}\}
\end{equation}
with $\overline{P_b}=3.69$ and $\sigma_{\log P_b}=1.3$ for M-type primary stars and $\overline{P_b}=5.03$ and $\sigma_{\log P_b}=2.28$ for FGK-type primary stars. We use $m_\ell = 0.5$\,M$_\sun$ as a threshold value between M-type stars and stars of earlier type. In both cases we limit our draws in the $[0.0-10.0]$ $\log P_b$ interval. The lower limit allows us to avoid Roche lobe overflows, while the upper limit is dictated by observations \citep[e.g.][]{Hartman20} which suggest that there is no clear cut in the long period tail of the distribution. After the generation of the mass of secondary star, which will be described in the next subsection, we use $P_b$ to calculate the binary semi-major axis $a_b$ using Newton's version of the third Kepler law.

\subsubsection{The mass ratio distribution}

As a third step, we  draw a value of mass ratio of the two stars of the system, $q=m_B/m_A$, from a probability density distribution, $\xi(q)$. We explore three possibilities, namely:
\begin{enumerate}
\item a single slope power-law distribution, as reported by \citet[][hereafter DK13]{DucheneKraus13} in their table 1;
\item a log-normal distribution, as suggested by \citet[][hereafter DM91]{Duquennoy91};
\item a double slope power-law with an excess fraction of same mass stars, as suggested by \citet[][hereafter MDS17]{Moe17}.
\end{enumerate}
In the first case, the distribution will have the form:
\begin{equation}
\xi (q) \propto q^\gamma
\end{equation}
with $\gamma=0.4$ for M-type stars and $\gamma=0.3$ for FGK-stars, using again $m_\ell = 0.5$\,M$_\sun$ as a conventional limit between M-type and FGK-type stars. This type of distribution favors larger values of $q$.

In the second case, the distribution will have the form:
\begin{equation}
\xi(q) \propto \exp [ -(q-\overline{q})/2 \sigma_q^2]
\end{equation}
with $\overline{q}=0.23$ and $\sigma_q=0.42$ over the entire mass range. This type of distribution favors average-to-low values for $q$.

In the third case, the distribution will have the form:
\begin{equation}
\label{eq.q_mds17}
\xi(q) \propto
\begin{cases}
q^{\gamma_L}            & \text{if}~ q<0.3      \\
q^{\gamma_H}            & \text{if}~ 0.3<q<0.95 \\
q^{\gamma_H}+\text{Exc} & \text{if}~ q>0.95     \\
\end{cases}
\end{equation}
where $\gamma_L$ and $\gamma_H$ are the slopes for low and high values of $q$, respectively, and \text{Exc} is the excess fraction of systems with same-mass stars, aka twins. This distribution is valid for F- and G-type stars, while for K- and M-type stars we will revert to DM91. For the slopes of the power law part of the distribution, $\xi_\text{pwl}(q)$, we have:
\begin{equation}
\gamma_L = 0.3 
\end{equation}
\begin{equation}
\gamma_H = 
\begin{cases}
-0.5                  & \text{if}~ \log P_b  <5   \\
-0.5-0.3(\log P_b -5) & \text{if}~ \log P_b \ge 5 \\
\end{cases}
\end{equation}
On the other hand, the excess fraction will be calculated as follows:
\begin{equation}
\label{eq.twin0}
F_\text{twin,0}= 0.3 - 0.15\log \biggl(\frac{m}{M_\sun}\biggl)
\end{equation}
\begin{equation}
\label{eq.twin}
F_\text{twin}=
\begin{cases}
F_\text{twin,0} & \text{if}~ \log P_b<1 \\
F_\text{twin,0}\bigl(1-\frac{\log P_b-1}{7}\bigl) & \text{if}~ 1<\log P_b<8\\
0 & \text{if}~ \log P_b>8
\end{cases}
\end{equation}
\begin{equation}
\label{eq.exc}
\text{Exc}=\frac{F_\text{twin}\int_{0.3}^{1.0}\xi_\text{pwl}(q)dq}{1-F_\text{twin}}
\end{equation}
Eq.~\eqref{eq.twin0} gives the excess fraction (as intended in MDS17, see their Fig. 2) for short period binary systems and Eq.~\eqref{eq.twin} scales this excess for higher periods. Finally, Eq.~\eqref{eq.exc} gives the actual excess fraction of twins (i.e. the red area in Fig. 2 of MDS17) to be inserted in Eq.~\eqref{eq.q_mds17}. It should be noted that we are extending to $\log P=10$ the results of MDS17, that they derived for systems with periods up to $\log P=8$. When $\log P_b<5$ this distribution has two bumps, namely at $q=0.3$ and at $q=1.0$, and is flat overall. On the other hand, when $\log P_b>5$ it becomes increasingly skewed towards the first bump and for high $P_b$ is similar to the DM91 case, favoring average-to-low values of $q$.

After normalizing the distribution ($\int_0^1 \xi(q)\,dq=1$), we use $\xi(q)$ to draw a value of $q$ and calculate the mass of the secondary star as $m_\text{B}=q\,m_\text{A}$ from the value of $m_\text{A}$ drawn in the first step (see \ref{sectPrimaryMass}). Since $q$ has values in the $[0-1]$ range, $m_\text{B}$ is guaranteed to be less than or equal to $m_\text{A}$.

\subsubsection{The binary eccentricity distribution}

We finally assign the orbital eccentricity, $e_b$, using another empirical distribution, $\xi(e_b)$. The eccentricity distribution is strongly affected by the orbital period: for $P_b$ less than a certain value identified as the circularization period $P_\mathrm{circ}$ the orbits are circularized due to the long-term history of tidal interactions between the two stars \citep{Duquennoy91}. On the other hand, the eccentricity distribution shows a remarkably small dependence on the mass of the primary \citep{DucheneKraus13}. For our calculation, we test two different hypotheses:
\begin{enumerate}
\item a Gaussian distribution with $P_\mathrm{circ} = 20$ days ($\log P_b=1.3$), as suggested by \citet[][hereafter SB01]{StepinskiBlack01};
\item a period-dependent power law distribution as suggested by MDS17.
\end{enumerate}

In the first case, the function will have the form:
\begin{equation}
\xi(e_b) \propto \exp [ -(e_b-\overline{e})^2/2 \sigma_e^2]
\end{equation}
$\overline{e}= 0.35$ and $\sigma_e=0.2$ for all values of $P_b$ above 20 days. This produces a sample with low-to-average eccentricities.

In the second case, the function will have the form:
\begin{equation}
\xi(e_b) \propto e^\eta \quad \text{with} \quad \eta=0.6-\frac{0.7}{\log P_b-0.5}
\label{eqMDS17}
\end{equation}
This power law favors low eccentricities when $\log P_b < 5/3$ and favors high eccentricities when $\log P_b > 5/3$. Given that the distribution of periods produces far more binaries with $\log P_b > 5/3$, the average eccentricity of the sample will be skewed towards larger values than in the SB01 case. The circularization period adopted by MDS17 is $\log P_\text{circ}=0.3$, but their function is actually not normalizable if $\log P_b \le 0.9375$. We therefore adopted a circularization period equal to $\log P_\text{circ} = 0.94$, after testing that the final distribution of eccentricities is almost unaffected by this change. In fact, the fraction of systems generated with an eccentricity in the $[0.0-0.1]$ range is over 98\% when $\log P_b=0.94$, meaning that our results provide a good match to the observational data employed by MDS17 in their analysis. It should also be considered that MDS17 define a period-dependent maximum eccentricity, equal to:
\begin{equation}
e_\text{max}=1-\biggl(\frac{P_b}{2 ~\text{days}} \biggl)^{-2/3}
\end{equation}
that further reduces the cases with high eccentricity when the orbital period is low, making even less critical our choice $\log P_\text{circ} = 0.94$. As far as the upper end of the distribution is concerned, in our calculations we extrapolated Eq.~\eqref{eqMDS17} up to $\log P_b = 10$, that is the upper limit of period as described in sect.~\ref{sectPeriods}.

\subsection{Boundaries of dynamical stability}

The parameters $m_\text{A}$, $\mu=m_\text{B}/(m_\text{A}+m_\text{B})$, $a_b$, and $e_b$, drawn for each binary systems are used to explore which regions of parameter space are consistent with the existence of stable planetary orbits. To this end, we use the results of the study of orbital stability of planets in binary systems performed by \citet{Quarles18} and \citet{Quarles20}. These authors derived empirical expressions for calculating the largest circumstellar orbit and the smallest circumbinary orbit where fully-interacting particles survive the length of the integration ($10^5$ binary periods). The semi-major axis of such critical planetary orbits are approximated as functions of $e_b$ and $\mu$. In units of the binary semi-major axis, the semi-major axis of the smallest, stable P-type orbit is:
\begin{equation}
\frac{a_\text{crit,P}}{a_b}=1.48+3.92 \, e_b - 1.41 \, e_b^2 + 5.14 \, \mu  +0.33 \, e_b \, \mu  -7.95 \, \mu^2 -4.89 \, e_b^2 \mu^2  
\label{acritP}
\end{equation}
whereas the semi-major axis of the largest, stable S-type orbit is:
\begin{equation}
\frac{a_\text{crit,S}}{a_b}=0.501-0.435 \, \mu -0.668 \, e_b +0.644 \, \mu \, e_b + 0.152 \, e_b^2 - 0.196 \, \mu \, e_b^2 ~~.
\label{acritS}
\end{equation}
This last expression is used  for the circumstellar stability regions of both stars. For the primary, $a_\text{crit,SA}$ is estimated adopting $\mu=m_\text{B}/(m_\text{A}+m_\text{B})$, whereas for the secondary $a_\text{crit,SB}$ is estimated adopting $\mu=m_\text{A}/(m_\text{A}+m_\text{B})$. We use the above empirical expressions within the limits of validity given by  \citet{Quarles18,Quarles20}, namely $e_b \leq 0.8$ and $\mu \in [0.01,0.99]$.

\subsection{Luminosity and effective temperature of the stars}

After extracting the masses of the primary ($m_\text{A}$) and secondary star ($m_\text{B}$), the luminosities ($L_\text{A}$ and $L_\text{B}$) and effective temperatures ($T_\text{A}$ and $T_\text{B}$) are calculated using the PARSEC evolutionary tracks of main sequence stars by \citet{Bressan12}. The main sequence is the most interesting stage for planetary habitability thanks to the extremely low rate of luminosity evolution.
For a given evolutionary track we use the mid point of the main sequence to assign the effective temperature and luminosity from a given stellar mass. We consider two different sets of tracks, one with solar metallicity ($Z=0.017$) and one with subsolar metallicity ($Z=0.008$), representative of stars in earlier stages of Galactic chemical evolution. To avoid extrapolation of stellar evolutionary tracks, we exclude cases with $m_\text{B} < 0.08$ M$_\sun$, the mass threshold limit for hydrogen burning.

\subsection{Boundaries of insolation}
\label{sect_radiative_Boundaries}

As in the case of  single stars, the insolation boundaries of the circumbinary and circumstellar habitable zones of binaries systems depend not only on the integrated stellar flux, but also on the spectral energy distribution of the stellar radiation. The reason for this is that hotter stars emit a larger fraction of energy in the form of short-wavelength radiation that is reflected back to space from the planetary atmosphere via Rayleigh scattering. As a result, the Bond albedo of the planet increases with the effective temperature of the star \citep{Selsis07,Kopparapu13a}. This effect can be taken into account in binary systems by summing the fluxes of the two stars with spectral weight factors that take into account the SED of the star and the radiative properties of the planetary atmosphere \citep{Kaltenegger13,Haghighipour13}. Following this approach we can estimate the combined flux of the two stars with the expression 
\begin{equation}
\label{eff_flux}
F_\text{eff,x}= w(T_1,\mathcal{R}_x)\frac{L_1}{r^2_\text{p-1}}+w(T_2,\mathcal{R}_x)\frac{L_2}{r^2_\text{p-2}}  ~~~,
\end{equation}
where $L_i$  is the luminosity of each of the two stars ($i=1,2$) and $r_\text{p-$i$}$ the distance between the planet and each star. The term $w(T_i,\mathcal{R})$ is a spectral weight dependent on the effective temperature of each star, $T_i$, and on a set of atmospheric parameters, $\mathcal{R}_x$, that specify the radiative properties of the planetary atmosphere (e.g. clouds, atmospheric composition, stratification, surface pressure, etc.). 

To define the insolation boundaries in a binary system, we calculate at which distance from  the stars, $r_\text{p-1}$ and  $r_\text{p-2}$, the effective flux \ref{eff_flux} matches the flux received by a planet at the inner and outer edge  (x = in, out) of the classic HZ around a single star \citep{Kasting93}. In practice, we impose the condition: $F_\text{eff,x}=L_\text{Sun}/l^2_\text{x-Sun}$, where $l_\text{x-Sun}$ is the inner or outer edge of the HZ around the Sun. We adopt the parametrization of $l_\text{x-Sun}$ limits provided by \citet{Kopparapu13a,Kopparapu13b}. These limits are calculated with single-column atmospheric models considering Runaway Greenhouse (RG) conditions  for the inner edge and Maximum Greenhouse for the outer edge. The radiative atmospheric parameters, $\mathcal{R}_\text{x}$, mimic an H$_2$O-dominated atmosphere at the inner edge and a CO$_2$-dominated atmosphere at the outer edge. In addition, two empirical limits are considered, namely the Recent Venus (RV) and Early Mars (EM), inferred from the possible existence of liquid water on Venus (up to $1$ Gyr ago) and Mars (up to $3.5$ Gyr ago), respectively.
The flux limits for these four edges, taken from \citet{Kopparapu13b}, are listed in the first row of Table \ref{coeffs}. The corresponding values of  $l_\text{x-Sun}$ (AU) that we adopt are listed in the second row of the same table.
As far as the spectral weights are concerned, following \citet{Kaltenegger13}, we adopt the equation
\begin{equation}
\label{eqweights}
w(T_i,\mathcal{R}_\text{x})=[1+\alpha_\text{x}(T_i,\mathcal{R}_\text{x}) \, l^2_\text{x-Sun}]^{-1}
\end{equation}
where
\begin{equation}
\label{alphas}
\alpha_\text{x}(T_i,\mathcal{R}_\text{x})=\text{a}_\text{x}\tilde{T}_i+\text{b}_\text{x}\tilde{T}^2_i+\text{c}_\text{x}\tilde{T}^3_i+\text{d}_\text{x}\tilde{T}^4_i
\end{equation}
with $\tilde{T}_i=T_i-5780$ K. The term $\alpha_\text{x}(T_i,\mathcal{R}_\text{x})$ captures the dependence of the planetary Bond albedo on the stellar effective temperature for a given set of atmospheric planetary conditions, $\mathcal{R}_\text{x}$. Higher stellar effective temperatures give weights \eqref{eqweights} smaller than 1, while for lower stellar effective temperatures the opposite is true. The coefficients $\text{a}_\text{x}$, $\text{b}_\text{x}$, $\text{c}_\text{x}$, $\text{d}_\text{x}$, derived from a 4-th order fit in $T_i$ \citep{Kopparapu13a,Kopparapu13b}, are also shown in Table \ref{coeffs}. These coefficients are valid for main sequence stars in the range of effective temperature $2600<T_i <7300$ K.
 
\begin{table}
\begin{center}
\begin{tabular}{lcccc}
\hline
Constant & Runaway & Maximum & Recent & Early \\
value & Greenhouse & Greenhouse & Venus & Mars \\
\hline
$F_\text{x}$  ($\text{S}_\earth$)    & 1.0385 & 0.3507 & 1.7763 & 0.3207 \\
$l_\text{x-Sun}$ (AU)& 0.9813 & 1.6882 & 0.7503 & 1.7658 \\
a$_\text{x}$ &  $1.2456\times10^{-4}$  &  $5.9578\times10^{-5}$  &  $1.4335\times10^{-4}$  &  $5.4471\times10^{-5}$ \\
b$_\text{x}$ &  $1.4612\times10^{-8}$  &  $1.6707\times10^{-9}$  &  $3.3954\times10^{-9}$  &  $1.5275\times10^{-9}$ \\
c$_\text{x}$ & $-7.6345\times10^{-12}$ & $-3.0058\times10^{-12}$ & $-7.6364\times10^{-12}$ & $-2.1709\times10^{-12}$\\
d$_\text{x}$ & $-1.7511\times10^{-15}$ & $-5.1925\times10^{-16}$ & $-1.1950\times10^{-15}$ & $-3.8282\times10^{-16}$\\
\hline
\end{tabular}
\end{center}
\caption{Flux limits and coefficients of Eq. \eqref{alphas} from  \citep{Kopparapu13b}. Fluxes are in units of the mean orbital Earth insolation, $\text{S}_\earth$.}
\label{coeffs}
\end{table}

To derive the circumbinary and circumstellar boundaries of insolation we consider circular planetary orbits with constant orbital radius $r_\text{p-1}=a_1$ around the host star (hereafter ``star 1''). The case of eccentric planetary orbits is discussed later (Section \ref{sectTrendPlanEcc}). We limit our attention to planetary orbits that lie inside, or in the proximity, of the regions of dynamical stability.

For circumbinary (P-type) orbits the critical semi-major axis of dynamical stability, $a_\text{crit,P}$, is larger than the typical distance between the two stars,  $r_{12} \simeq a_b$. Since stable orbits have $a_1 > a_\text{crit,P}$, we assume that in the circumbinary  region of dynamical stability the planet will be sufficiently distant from from both stars to justify the approximation $r_\text{p-2} \simeq r_\text{p-1}$. With this approximation the condition $F_\text{eff,x}=L_\text{Sun}/l^2_\text{x-Sun}$ yields:
\begin{equation}
w(T_1,\mathcal{R}_\text{x})\frac{L_1}{a_\text{x,cb}^2}+w(T_2,\mathcal{R}_\text{x})\frac{L_2}{a_\text{x,cb}^2} =\frac{L_\text{Sun}}{l^2_\text{x-Sun}} ~~,
\label{eqPorbitEdge}
\end{equation} 
where $a_\text{x,cb}$ is the inner or outer edge of the radiative HZ in the circumbinary region. From this we obtain:
\begin{equation}
a_\text{x,cb}=\left[ {  w(T_1,\mathcal{R}_\text{x}) \, \widetilde{L}_1 +w(T_2,\mathcal{R}_\text{x})) \, \widetilde{L}_2  } \right]^{1/2} l_\text{x-Sun}
\label{eqPedge}
\end{equation}
where $\widetilde{L}_i=L_i/L_\text{Sun}$. This equation is valid if $a_\text{x,cb}$ is sufficiently close to the circumbinary region of dynamical stability so that $a_\text{crit,P}/a_b \gg 1$ and $r_\text{p-2} \simeq r_\text{p-1}$. From our simulations we find that this is generally true.

For circumstellar (S-type) orbits, the mean orbital insolation from the host star is constant, but the flux received from the other star (hereafter ``star 2'') evolves in time according to the instantaneous $r_\text{p-2}=r_\text{p-2}(t)$. In a reference frame centered in star 1, we have:
\begin{equation}
\label{eqrp2}
r_\text{p-2}=[r^2_\text{p-1} + r^2_\text{12} + 2r_\text{p-1}r_\text{12}(\nu)\cos{\theta} ]^{1/2}
\end{equation}
where $\nu$ is the true anomaly (i.e. the position angle of star 2 along the stellar orbit with respect of the periastron) and $\theta$ is the position angle of the planet orbiting star 1. A sketch of this astrocentric configuration can be seen in the bottom panel of fig.1 in \citep{Kaltenegger13}. To derive $r_\text{p-2}(t)$ we express $\theta$ and $\nu$ as functions of $t$. For circular orbits around the host star, $\theta=(2\pi/P_p)t+\theta_0$, where $P_p$ is the orbital period of the planet. On the other hand, $\nu$ does not increase linearly with time. To solve this problem, we replace the true anomaly with eccentric anomaly, $E$. This angle can be calculated using Kepler's equation
\begin{equation}
\label{eqEcc}
E(t) - e_b \sin{E(t)} = M(t) \quad ,
\end{equation}
where the mean anomaly $M(t)=(2\pi/P_b)t + M_0$, increases linearly with time and $P_b$ is the orbital period of the star 2. With this formalism the distance between the two stars is
\begin{equation}
\label{eqR12}
r_{12}(t)=a_b [1 - e_b \cos{E(t)}] \quad .
\end{equation}
We solve this equation numerically\footnote[5]{See \citet[][appendix A.5]{Vladilo15} for details on the numerical solution of Eqs.~\eqref{eqEcc} and \eqref{eqR12}} and insert the result in Eq.~\eqref{eqrp2} to obtain $r_\text{p-2}(t)$. Having derived $r_\text{p-2}(t)$, we use Eq.~\eqref{eff_flux} to calculate the instantaneous value of the circumstellar radiative boundary, $a_\text{x,cs1}$, that satisfies the condition $F_\text{eff,x}=L_\text{Sun}/l^2_\text{x-Sun}$. Given the dependence of $r_\text{p-2}$ on $r_\text{p-1}$, shown in \eqref{eqrp2}, the solution is found using numerical methods. The results are strongly dependent on the eccentricity $e_b$ and the semi-major axis $a_b$ of the binary system. When $e_b$ is high and $a_b$ is small, the boundary varies dramatically in time and it is meaningless to derive a representative value of the circumstellar radiative boundary. Luckily, these cases are automatically excluded from our analysis of habitability because, when such conditions hold, $a_\text{x,cs1}(t)$ overlaps invariably the region of dynamical instability. In practice, when $a_\text{x,cs1}(t) \ge a_\text{crit,S1}$ during portion of the orbital period, we tag the case as dynamically unstable. Instead, for the majority of systems extracted, we find $a_\text{x,cs1}(t) < a_\text{crit,S1}$ during the whole orbital period. When this condition holds, the relative contribution of star 2 to the total insolation evolves smoothly or is negligible, because $e_b$ is small or/and $a_b$ is large. In this case we derive a mean value of $a_\text{x,cs1}(t)$ by averaging the time-dependent terms of Eq.~\eqref{eff_flux}. This approach is sufficient for our purpose, given the statistical nature of our calculations. From the condition $F_\text{eff,x}=L_\text{Sun}/l^2_\text{x-Sun}$ we obtain
\begin{equation}
\label{eq_numflux}
w(T_1,\mathcal{R}_x) \frac{L_1}{r^2_\text{p-1}} + w(T_1,\mathcal{R}_x) \frac{L_1}{\Delta t} \int_0^{\Delta t} \frac{dt}{r^2_\text{p-2}(t)} = \frac{L_\text{Sun}}{l^2_\text{Sun}} \quad .
\end{equation}
We solve this equation numerically in the unknown $r_\text{p-1}$, taking into account the dependence of $r_\text{p-2}$ on $r_\text{p-1}$. The solution $r_\text{p-1}=a_\text{x,cs1}$ is adopted as the circumstellar radiative boundary. In the numerical integration we adopt a variable time step sufficiently small to sample properly the planetary and stellar periods (i.e. $\delta t \ll \text{min}(P_b,P_p)$) and an integration interval $\Delta t$ sufficiently long to obtain a stable solution. In this way the estimate of $a_\text{x,cs1}$ is not affected by the choice of the initial conditions $\theta_0$ and $M_0$. As an additional test of dynamical stability, we checked if the planet enters the Hill radius of the star 2 in the course of the binary orbital evolution. This test shows that the Hill radius criterion is always less stringent than the criterion $a_\text{x,cs1} \ge a_\text{crit,S1}$. We therefore implemented only this last one, as explained above, to discriminate unstable cases for our analysis of habitability. Finally, in order to save computational time, when $a_b \gg a_p$ we approximate \eqref{eq_numflux} with the equation
\begin{equation}
\label{eq_anflux}
w(T_1,\mathcal{R}_x) \frac{L_1}{r^2_\text{p-1}} + w(T_1,\mathcal{R}_x) \frac{L_2}{a_b(1-e_b^2)^2} = \frac{L_\text{Sun}}{l^2_\text{Sun}}
\end{equation}
that we solve analytically. This approximation is justified   when $a_b \gg a_p$ because in this case the distance between the planet and star 2 is virtually identical to the distance between the two stars and the mean insolation from star 2 equals that received by a planet orbiting star 2 with semimajor axis $a_b$ and eccentricity $e_b$.\footnote{In our case the problem is symmetric, with a source in Keplerian orbit around the planet, but the mean orbital insolation \citep{Williams02} is obviously the same.} After some testing, we  found that $a_\text{crit,S1} > 10 \times a_\text{MG,cs1}$ is a good criterion for using Eq.~\eqref{eq_anflux} instead of Eq.~\eqref{eq_numflux}.

\begin{figure}[htbp!]
\plottwo{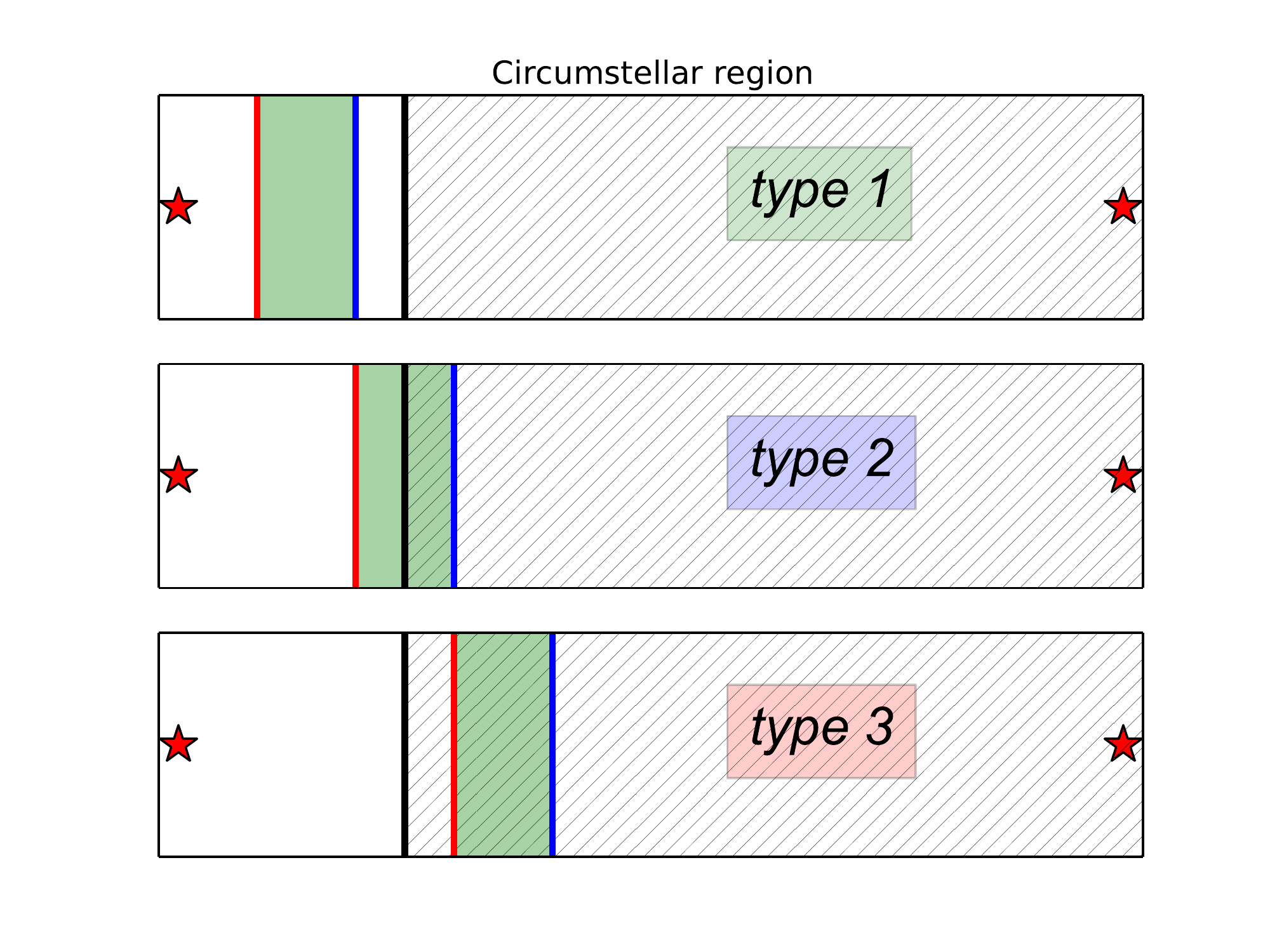}  {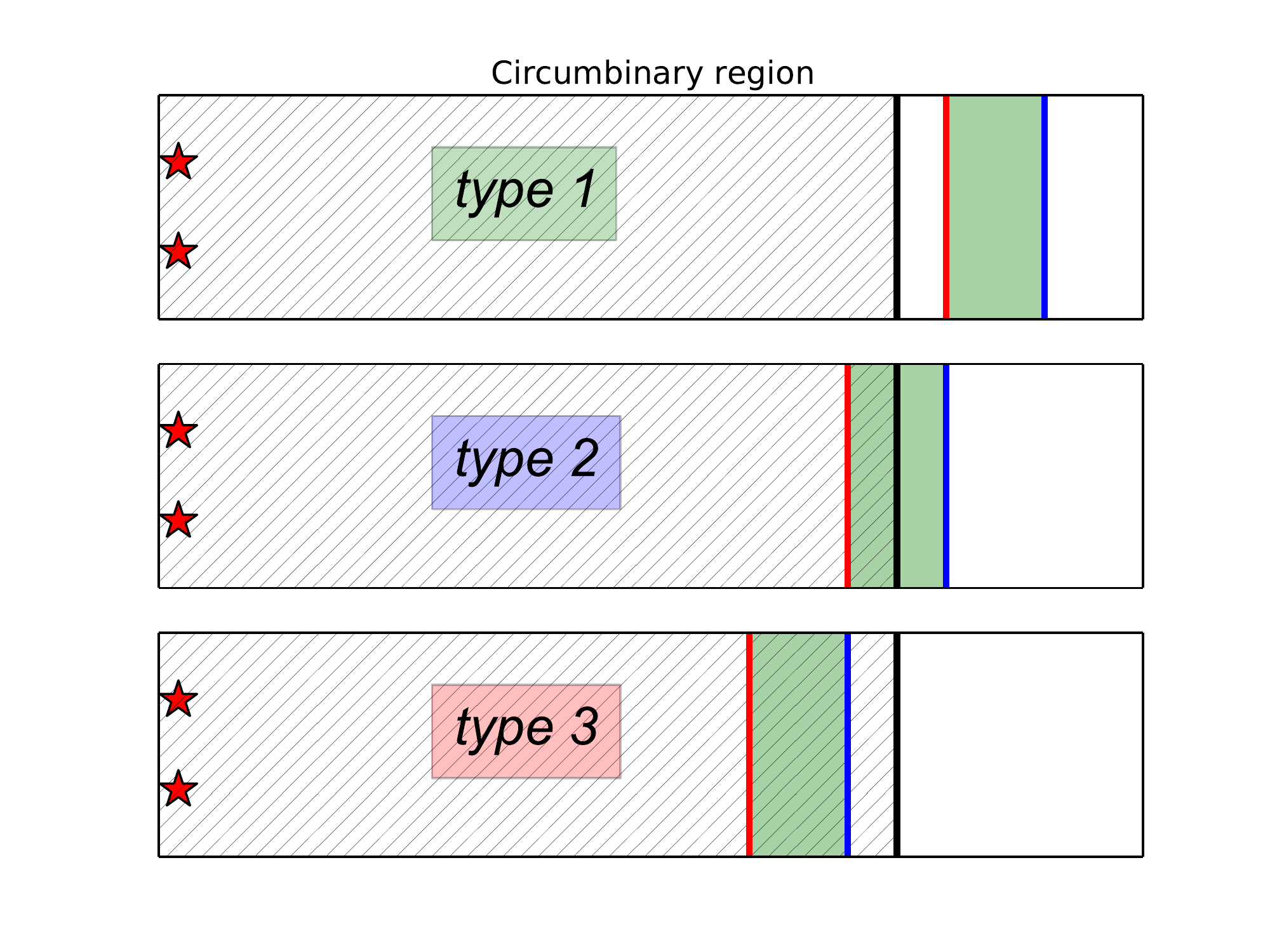}
\caption{Sketch of possible types of habitability conditions in binary systems. Left panels: circumstellar regions, with the host star to the left  and the companion star to the right. Right panels: circumbinary regions, with the pair of stars to the left. Green areas: radiative HZs. Red lines: inner edges of insolation, $a_\text{inn,cs}$ (left) and $a_\text{inn,cb}$ (right). Blue lines: outer edges of insolation, $a_\text{out,cs}$ (left) and $a_\text{out,cb}$ (right). Solid lines: boundaries of dynamical stability, $a_\text{crit,S}$ (left) and $a_\text{crit,P}$ (right); hatched areas: instability regions.
\label{HZtypes}}
\end{figure}

\begin{deluxetable}{cccccccccccc}[b!]
\tablecaption{Average statistical properties of habitability of the Monte Carlo samples of binary systems generated as explained in Section \ref{sectMonteCarlo}. The first 6 columns of the Table define the model that has been used to perform the simulations. The last 6 columns show the global indices of circumstellar and circumbinary habitability defined in Section \ref{sec:results}.}
\tablecolumns{13}
\tablehead{ \colhead{Model}  & \colhead{$\xi(\log m)\tablenotemark{a}$} & \colhead{$\xi(q)$} & \colhead{$\xi(e_b)$} & \colhead{$Z\tablenotemark{b}$}   &   \colhead{$e_p$} & \colhead{$f_\text{hab,SA}$} & \colhead{$\overline{w}_\text{SA}$} & \colhead{$f_\text{hab,SB}$} & \colhead{$\overline{w}_\text{SB}$} & \colhead{$f_\text{hab,P}$} & \colhead{$\overline{w}_\text{P}$}  }
\startdata
A & IMF & DK13 & SB01 & 0.017 & 0.0 & 0.843 & 0.955 & 0.869 & 0.970 & 0.035 & 0.930 \\
B & PDMF & DK13 & SB01 & 0.017 & 0.0 & 0.848 & 0.955 & 0.874 & 0.969 & 0.034 & 0.929 \\
C & IMF & DM91 & SB01 & 0.017 & 0.0 & 0.835 & 0.955 & 0.874 & 0.979 & 0.039 & 0.885 \\
D & PDMF & DM91 & SB01 & 0.017 & 0.0 & 0.839 & 0.955 & 0.877 & 0.977 & 0.036 & 0.882 \\
E & IMF & DK13 & SB01 & 0.008 & 0.0 & 0.832 & 0.954 & 0.859 & 0.969 & 0.038 & 0.941 \\
F & PDMF & DK13 & SB01 & 0.008 & 0.0 & 0.836 & 0.954 & 0.863 & 0.968 & 0.036 & 0.935 \\
G & IMF & DM91 & SB01 & 0.008 & 0.0 & 0.824 & 0.953 & 0.864 & 0.979 & 0.043 & 0.899 \\
H & PDMF & DM91 & SB01 & 0.008 & 0.0 & 0.829 & 0.952 & 0.869 & 0.977 & 0.039 & 0.888 \\
I & IMF & MDS17 & MDS17 & 0.017 & 0.0 & 0.792 & 0.947 & 0.838 & 0.969 & 0.045 & 0.918 \\
J & PDMF & MDS17 & MDS17 & 0.017 & 0.0 & 0.796 & 0.945 & 0.841 & 0.967 & 0.043 & 0.910 \\
K & IMF & MDS17 & MDS17 & 0.008 & 0.0 & 0.779 & 0.944 & 0.828 & 0.967 & 0.048 & 0.931 \\
L & PDMF & MDS17 & MDS17 & 0.008 & 0.0 & 0.783 & 0.943 & 0.831 & 0.966 & 0.045 & 0.925 \\
\enddata
\tablenotetext{a}{IMF: initial mass function; PDMF: present-day mass function (Chabrier 2003b).}
\tablenotetext{b}{Metallicity of the adopted stellar evolutionary track (Bressan et al. 2012).}
\label{tab:results}
\end{deluxetable}

\subsection{Binary habitable zones}
\label{BHZ}

By comparing dynamical boundaries with  insolation boundaries one can test if  binary systems have circumstellar and/or circumbinary zones that are habitable in the long term \citep{Cuntz14,Cuntz15}. In Fig.\,\ref{HZtypes} we sketch three types of habitability conditions than can be present in circumstellar and circumbinary regions (left and right panels, respectively). In what we call here ``type 1'' habitability, the whole radiative HZ is dynamically stable (top panels). Circumstellar and circumbinary ``type 1'' habitability is equivalent to Cuntz' ``S-type'' and ``P-type'', respectively. In ``type 2'', only part of the radiative HZ is dynamically stable (middle panels). Circumstellar and circumbinary ``type 2'' habitability is equivalent to Cuntz' ``ST-type'' and ``PT-type'', respectively. In ``type 3'', the region is uninhabitable because the whole radiative HZ is dynamically unstable (bottom panels). In accordance with Cuntz' analysis, we define the spatial extension of the circumbinary HZ as given by:
\begin{equation}
\Delta \ell_\text{P} = \max(a_\text{crit,P},a_\text{out,cb} )-\max(a_\text{crit,P},a_\text{inn,cb} ) ~~.
\label{DELTAhabP}
\end{equation}
When $\Delta \ell_\text{P} =0$, the binary system is uninhabitable for P-type orbits (type 3 condition, right panel of Fig. \ref{HZtypes}). This condition holds when $a_\text{crit,P} > a_\text{out,cb}$, in which case planets in dynamically stable  P-type orbits will be in a snowball state. Similarly, the spatial extension of the circumstellar HZ around each star is  given by:
\begin{equation}
\Delta \ell_{\text{S}\star} = \min(a_{\text{crit,S}\star},a_{\text{out,cs}\star} )-\min(a_{\text{crit,S}\star},a_{\text{inn,cs}\star} ) ~~,
\label{DELTAhabS}
\end{equation}
where $\star=$A and B for the primary and secondary, respectively. When $\Delta \ell_{\text{S}\star} =0$ the binary system is uninhabitable for S-type orbits around one of the two stars (type 3 condition, left panel of Fig. \ref{HZtypes}). This condition holds when $a_{\text{crit,S}\star} < a_{\text{inn,cs}\star}$, in which case planets in dynamically stable S-type orbits have an insolation above the runaway greenhouse limit.

It should be remarked that $\Delta \ell_{\text{S}\star}$ and $\Delta \ell_{\text{P}\star}$ vary with time according to the instantaneous location of the planet and the stars. This makes somewhat tricky to evaluate a continuously habitable zone. Where these variations are more pronounced, namely studying circumstellar cases in relatively close binaries, we calculate the time-averaged extension of the HZ (see Sect.~\ref{sect_radiative_Boundaries}). A part from these cases,  insolation variations due to stellar motions are either small enough to be basically negligible (e.g., for S-type HZs in wide binaries) or frequent enough to smooth out over orbital periods relevant for our problem (e.g. for all P-type HZs).

In the next section we present the statistical results that we obtain with our methodology, taking also into account our current understanding of planetary formation in binary systems. The interpretation of our results in light of observational surveys of exoplanets in binary systems is presented in Section \ref{sec:discussion}.

\section{Results} \label{sec:results}

To run the Monte Carlo simulations we considered different models, each model being defined by a specific  set of  distribution functions and stellar evolutionary tracks described in the previous section. The adopted prescriptions of the models are specified in the first 6 columns of Table \ref{tab:results}. For each model we performed a Monte Carlo extraction rejecting the cases with parameters outside the assigned ranges. Specifically, we rejected cases with mass of the secondary below the substellar threshold (0.08 $M_\sun$) and parameters $\mu$ and $e_b$ outside the range of validity of the prescriptions (\ref{acritP}) and (\ref{acritS}) given by  \citet{Quarles18,Quarles20}. The extraction procedure was iterated until an assigned number of systems, $N_\circ$, was generated. To assess the statistical robustness of the results, we compared the results obtained for increasing values of $N_\circ$, typically 10$^4$,  10$^5$,  and, for some models (namely A, B, C, D and F), $10^6$.

As a first step to study the results obtained from our procedure we introduced two indices aimed at characterizing the mean habitability of each sample. The first index, $f_\text{hab}$, is the fraction of systems, out of the total number $N_\circ$, for which the circumbinary or circumstellar regions are habitable. Specifically, the fraction of systems with habitable circumbinary regions, $f_\text{hab,P}$, is calculated counting the systems for which $\Delta \ell_\text{P}>0$. The fractions of systems with habitable circumstellar regions around the primary and the secondary, $f_\text{hab,SA}$ and $f_\text{hab,SB}$, are obtained counting the cases in which $\Delta \ell_\text{SA} >0$ and $\Delta \ell_\text{SB} >0$, respectively. 

The second index, $\overline{w}$, is a mean, normalized width of circumstellar or circumbinary HZs. We calculate this index normalizing the widths \eqref{DELTAhabP} and \eqref{DELTAhabS} to the width $\Delta \ell_\circ(m_1)$ of the HZ around a single star with mass $m_1$ (i.e. the mass of the host star). For circumbinary regions we set $m_1=m_A$ and calculate the circumbinary index $\overline{w}_\text{P}$ by averaging the ratio $w_\text{P}= \Delta \ell_\text{P}/\Delta \ell_\circ(m_A)$ for the systems for which $\Delta \ell_\text{P}>0$. Similarly, we calculate the circumstellar indices $\overline{w}_\text{SA}$ and $\overline{w}_\text{SB}$ by averaging the ratios $w_\text{SA} = \Delta \ell_\text{SA}/\Delta \ell_\circ(m_A)$ and $w_\text{SB} = \Delta \ell_\text{SB}/\Delta \ell_\circ(m_B)$ for the systems for which  $\Delta \ell_\text{SA} >0$ and $\Delta \ell_\text{SB} >0$, respectively. 

For both of these indexes, we found that the runs with $10^5$ extractions yield results that differ by less than 0.1\% compared to the test runs performed with  $10^6$ extractions. Also the distributions of different HZ types versus binary parameters are largely unaffected. Therefore we adopted the less-time consuming samples with $10^5$ to compare results obtained from different models. In the last 6 columns of Table \ref{tab:results} we list the indices $f_\text{hab}$ and $\overline{w}$ obtained by running a set of simulations with $N_\circ=10^5$ for each model. All calculations were performed using the climatological boundaries of habitability RG (runaway greenhouse) and MG (maximum greenhouse).

\begin{figure}[htbp!]
\plottwo{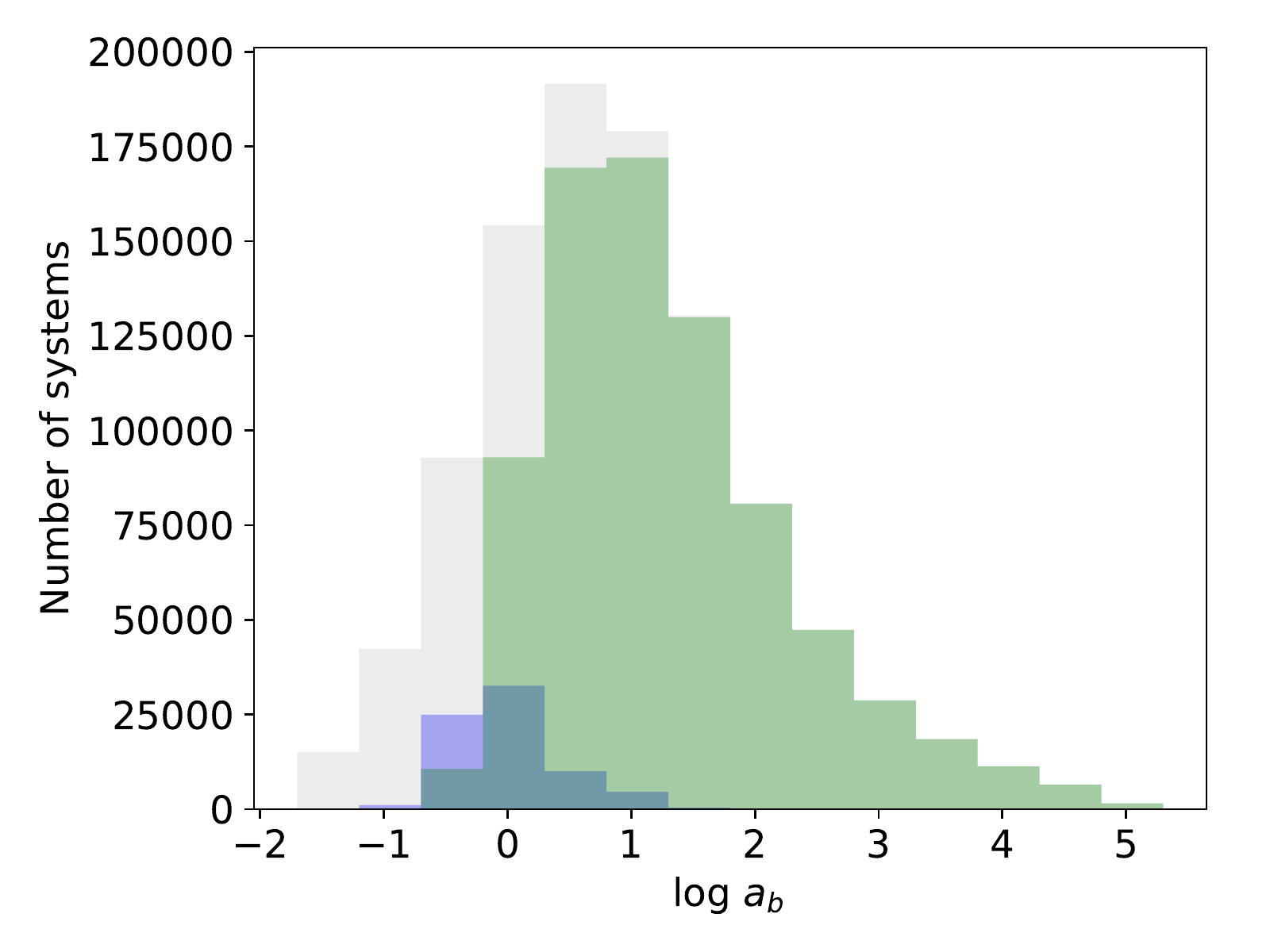}{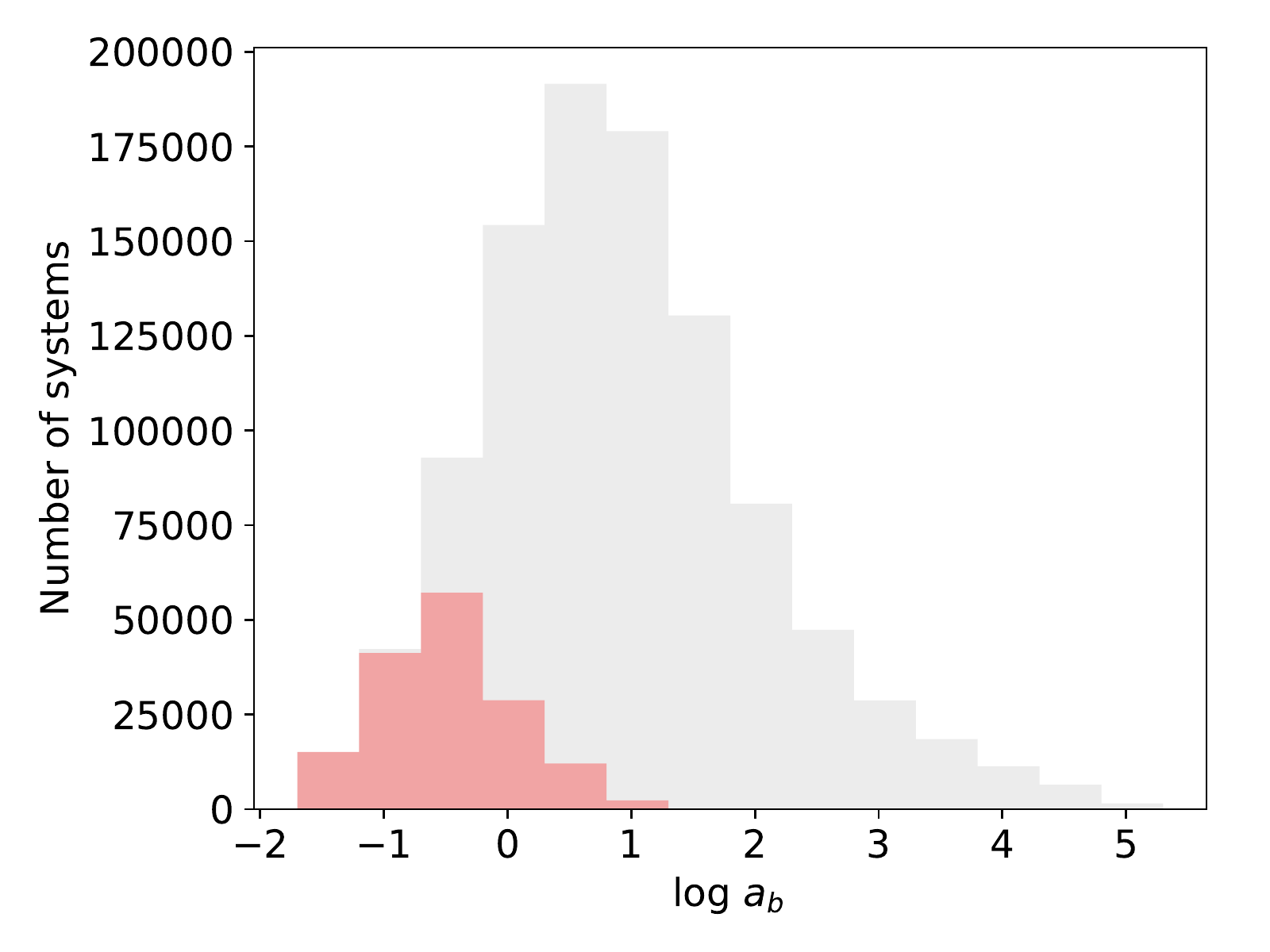}
\caption{Habitability of circumstellar regions around the primary star in binary systems. Gray histogram: number of binary systems  versus binary semi-major axis (AU) for the whole sample generated with Model A (Table \ref{tab:results}). Left panel: systems with $\Delta \ell_\text{SA}>0$ counted according to their conditions of habitability; green and blue histograms: type-1 and type-2 conditions. Red histogram in the right panel: systems with uninhabitable regions around the primary (type-3 condition). See Fig.\,\ref{HZtypes}. 
\label{fig_SA_ab_dist_hab}}
\end{figure}

\begin{figure}[htbp!]
\plottwo{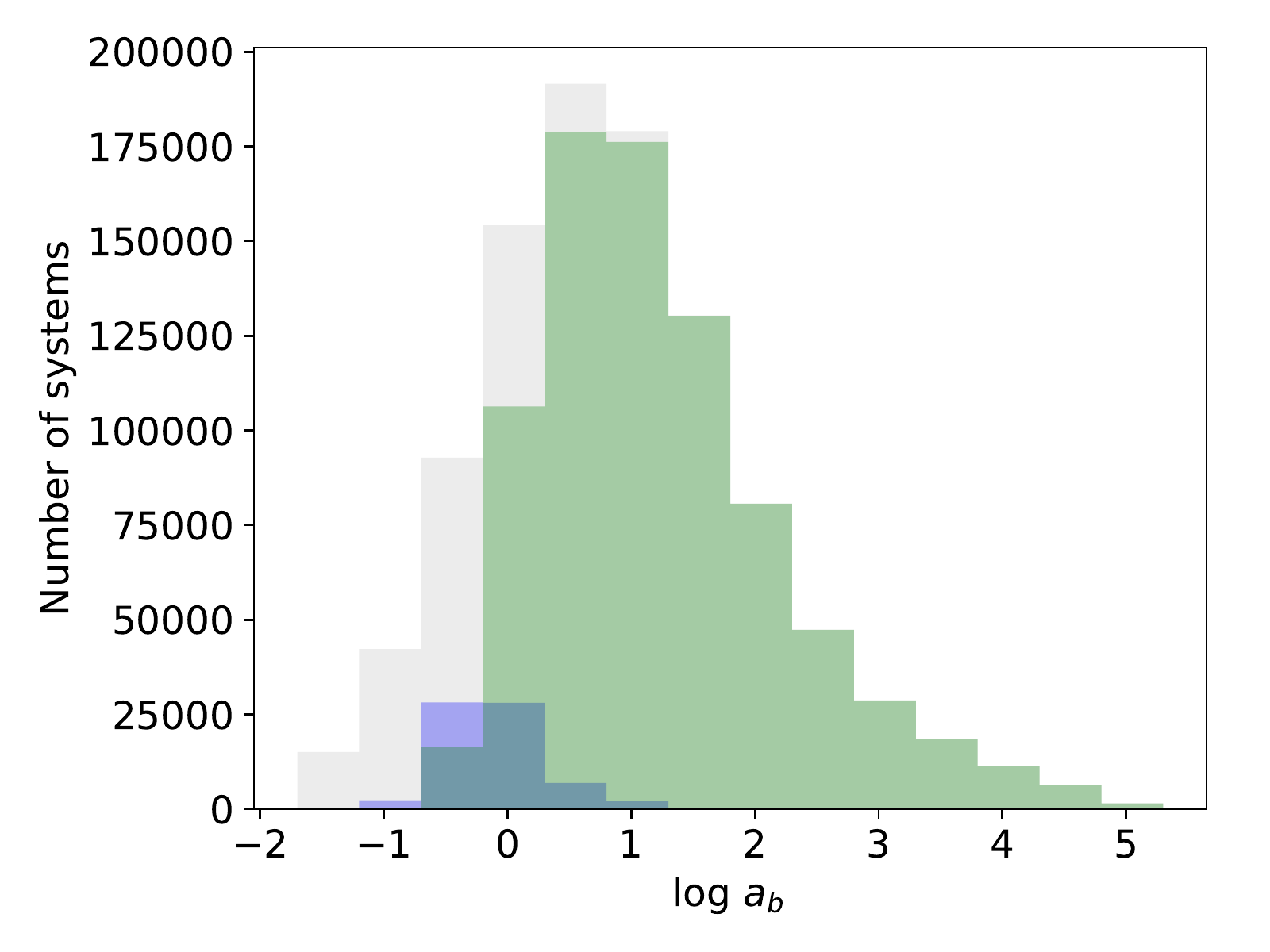}  {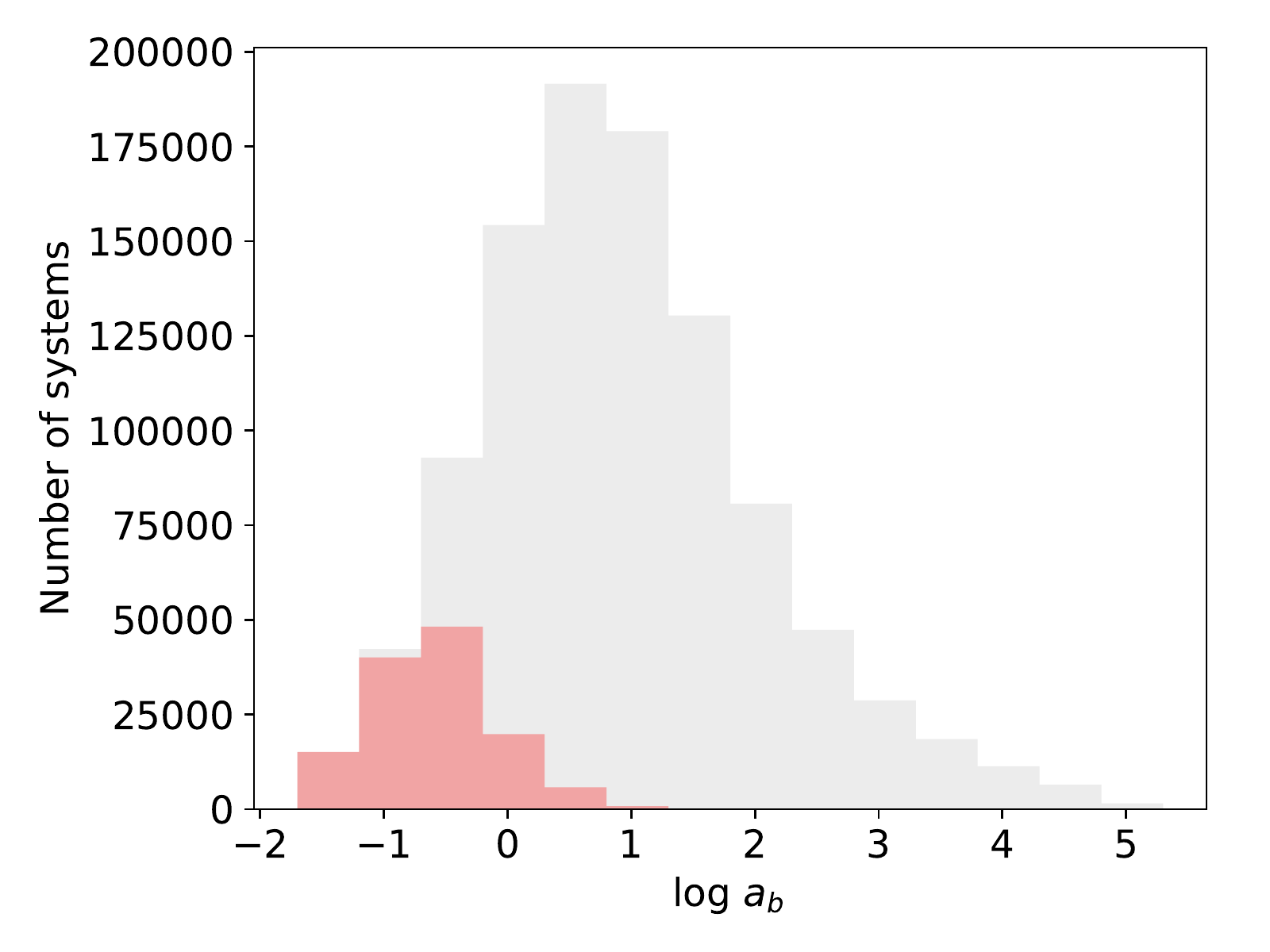}
\caption{Habitability of circumstellar regions around the secondary star in binary systems. Same sample of binary systems and same color coding of the histograms as in Fig. \ref{fig_SA_ab_dist_hab}.  
\label{fig_SB_ab_dist_hab}}
\end{figure}

\begin{figure}[htbp!]
\plottwo{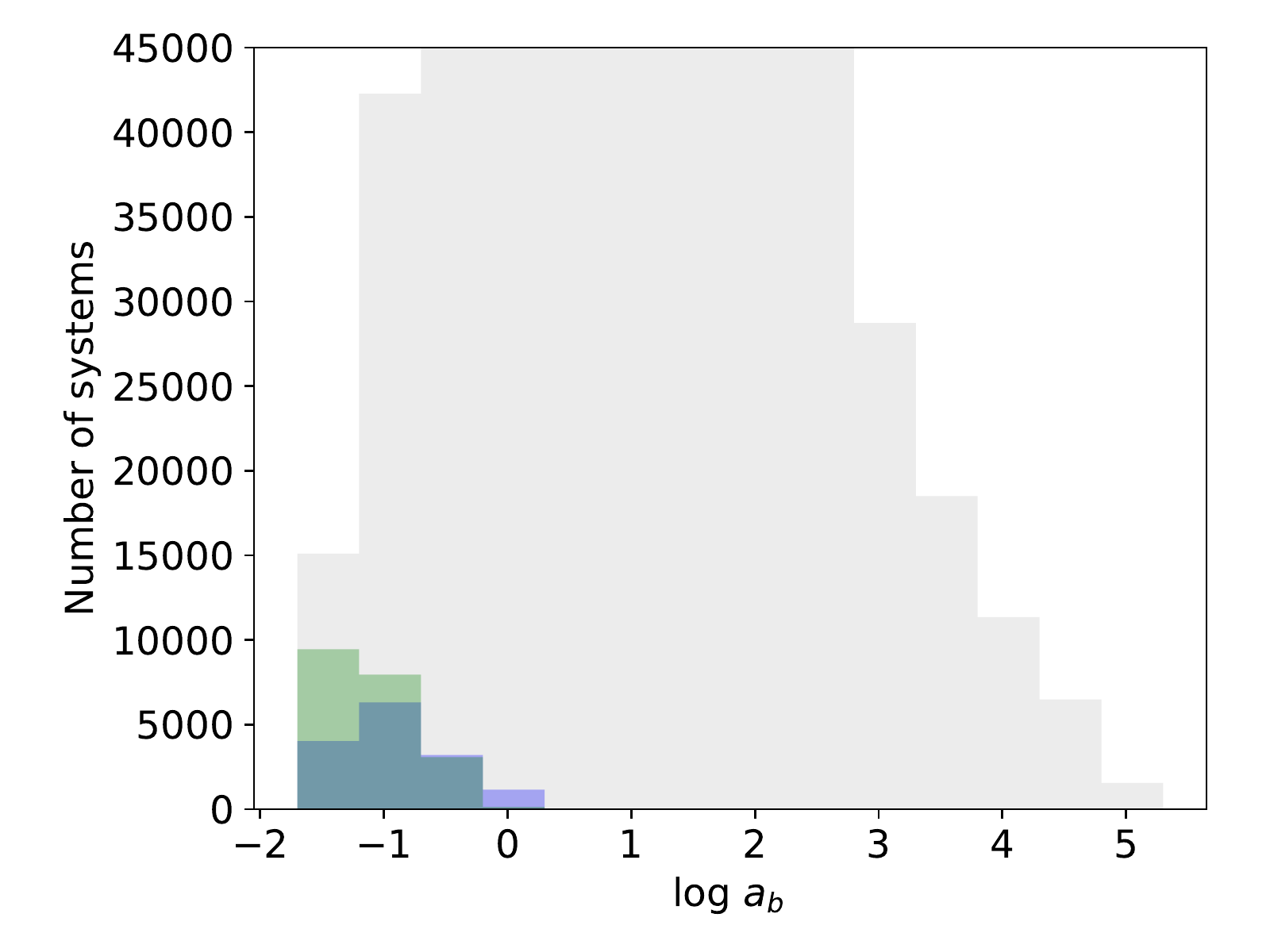}  {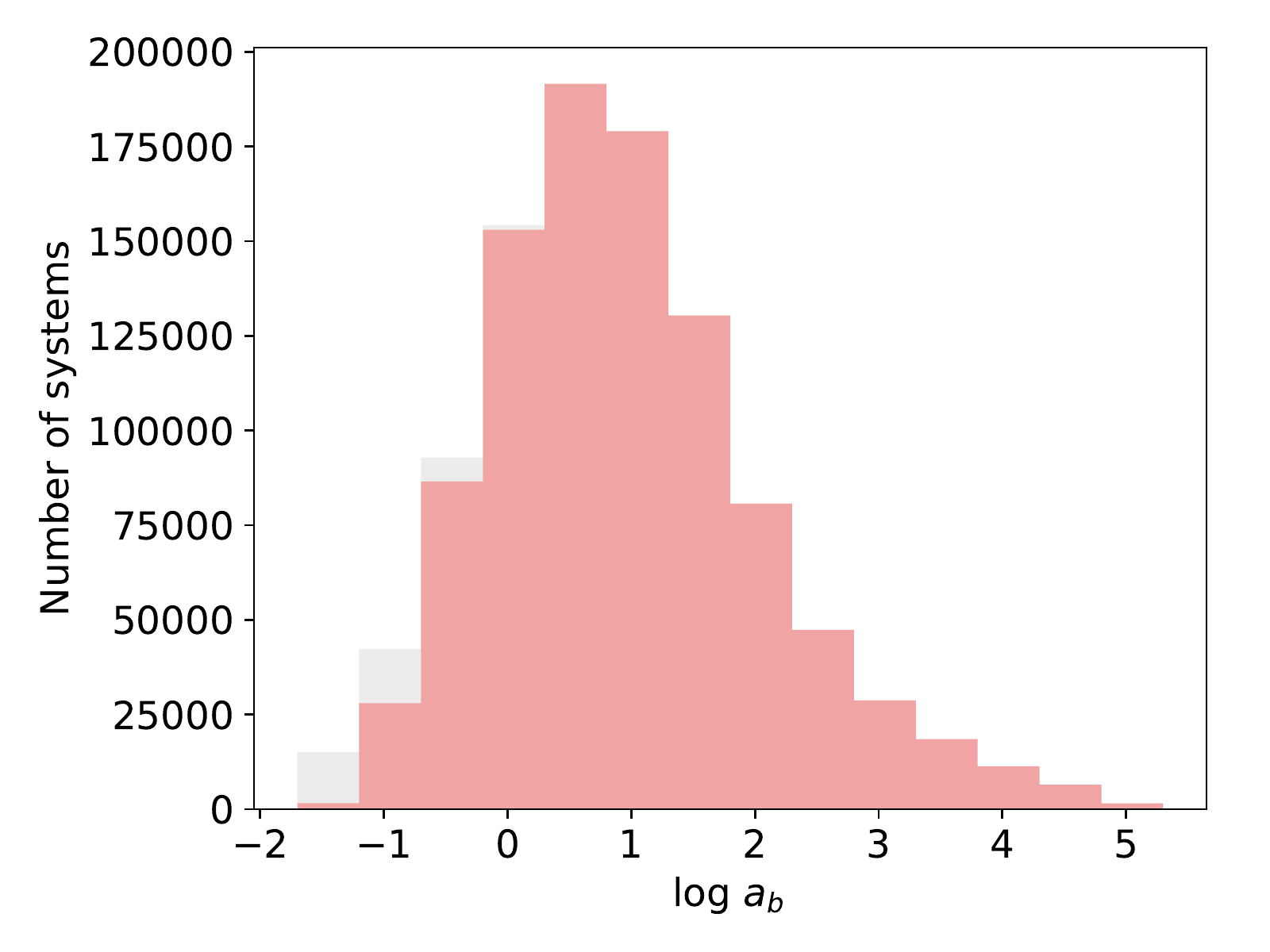}
\caption{Habitability of circumbinary regions in binary stellar systems. Gray histogram: whole sample of systems as in Fig. \ref{fig_SA_ab_dist_hab}. Left: expanded figure of systems with $\Delta \ell_\text{P}>0$. Right: systems with $\Delta \ell_\text{P}=0$. Same color coding of the histograms as in Fig. \ref{fig_SA_ab_dist_hab}. 
\label{fig_P_ab_dist_hab}}
\end{figure}

The results collected in Table \ref{tab:results} indicate that the average properties of habitability are very different for the circumstellar and the circumbinary regions. The main difference is that circumstellar regions are habitable in most systems, whereas circumbinary regions are uninhabitable in a large fraction of the full Monte Carlo sample. Specifically, for the circumstellar region around the primary all models yield $f_\text{hab,SA} \simeq 78-85\%$, with a mean, normalized width of the HZ, $\overline{w}_\text{SA} \simeq 0.95$. For the circumstellar region around the secondary, the habitable fraction is slightly higher, with $f_\text{hab,SB} \sim 84-88\%$ and the mean, normalized width is slightly larger, with $\overline{w}_\text{SB} \sim 0.97$. On the contrary, the circumbinary region has type-3 conditions in the majority of cases, yielding an average fraction of habitable circumbinary regions $f_\text{hab,P} \simeq 3-5\%$. However, for the circumbinary regions that are habitable, the mean, normalized width is quite high, $\overline{w}_\text{P} \sim 0.88-0.94$. In other words, circumbinary HZs are less numerous, but not less extended. As we discuss below, the fraction of habitable circumbinary regions is much higher in specific intervals of binary and stellar parameters. The fact that the mean widths of the HZs are almost equal to those around a single star implies that habitability conditions of type-2, where part of the radiative HZ is dynamically unstable, are not common. In fact, the detailed analysis that we present below shows that the HZs in binary systems can be
even more extended than in the case of a single star.

To cast light on the average statistical results summarized above, we investigated how the systems with type-1, type-2, and type-3 conditions of habitability are distributed  as a function of the binary semi-major axis. In Fig.\,\ref{fig_SA_ab_dist_hab} we show  the number of systems binned in constant logarithmic steps of $a_b$ for a sample obtained from model A (gray histogram) and for different sub-samples selected according to the habitability conditions around the primary star. In the left panel one can see that there is a majority of cases of type-1 habitability, with a maximum at $a_b \simeq 10$\,AU (green histogram), and a minority of cases with type-2 conditions, with a maximum at $a_b \simeq 1$\,AU (blue histogram). In the right panel one can see that the systems with uninhabitable circumprimary regions are  a minority, with a maximum at at $a_b \simeq 0.3$\,AU (red histogram). Similar results are found for the circumstellar regions around the secondary, shown in Fig. \ref{fig_SB_ab_dist_hab}. A close comparison with Fig. \ref{fig_SA_ab_dist_hab} shows that circumsecondary regions are slightly more habitable than circumprimary regions. This difference is not due to statistical fluctuations, but to intrinsically different conditions of habitability that we discuss below. 

The situation for circumbinary regions, shown in Fig.  \ref{fig_P_ab_dist_hab}, is completely different. In the right panel one can see that  the large majority of circumbinary regions is uninhabitable (type-3 conditions), with a maximum of cases at $\simeq 3$\,AU (red histogram). This maximum corresponds to the maximum of the period distribution of the entire sample. In the left panel we show the systems with type-1 (green histogram) and type-2 (blue histogram) conditions of habitability, where the former has a maximum for closest binary systems bin at $\simeq 0.02-0.06$\,AU and the latter has a maximum at $\simeq 0.1$\,AU. These findings were expected, since they generally reproduce the results of the seminal work on protoplanetary disks in binary systems by \citet{Arty94}. Results similar to those shown in Figs. \ref{fig_SA_ab_dist_hab}, \ref{fig_SB_ab_dist_hab} and \ref{fig_P_ab_dist_hab} are found for all models.

\begin{figure}[htbp!]
\plottwo{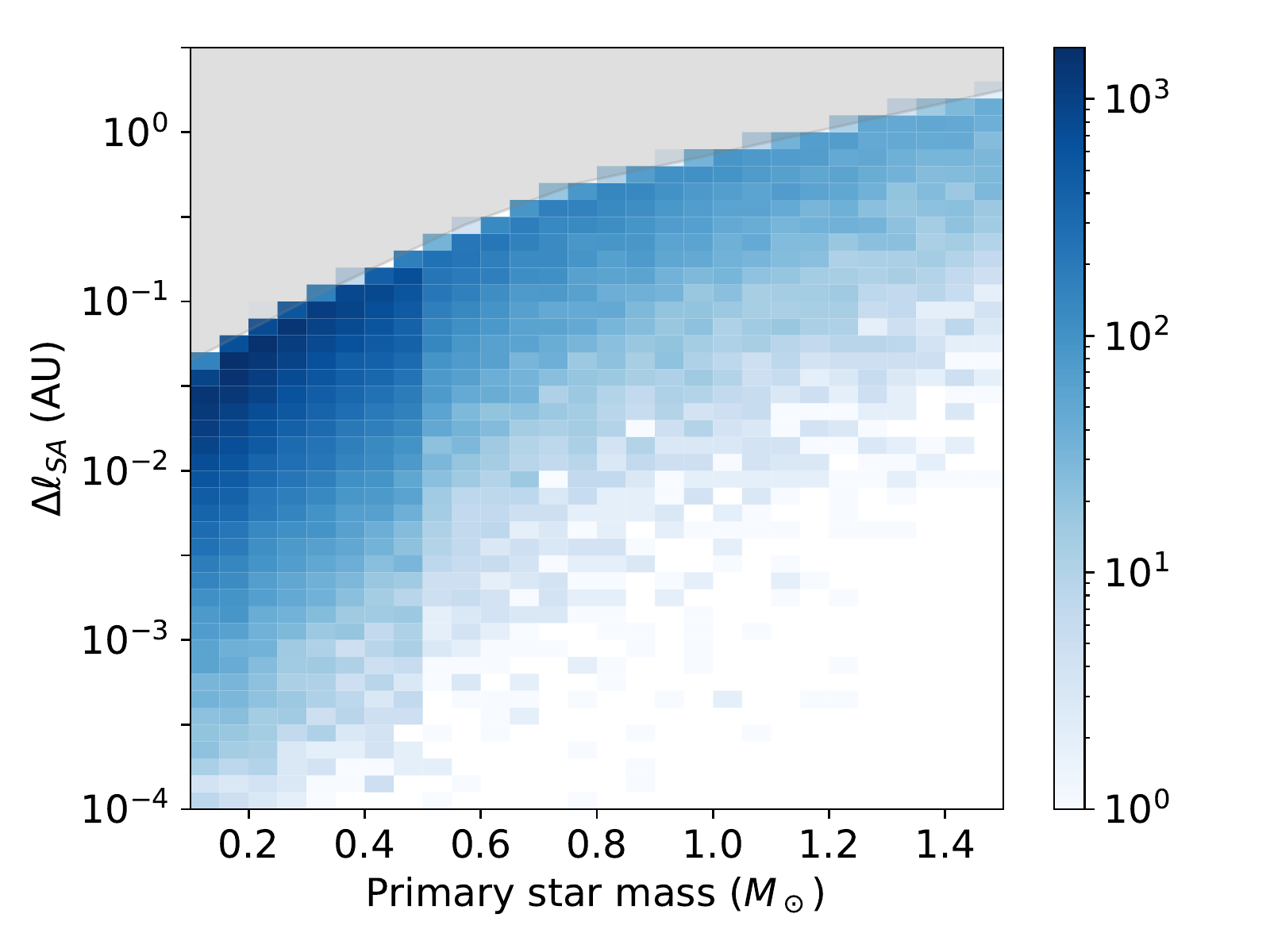}{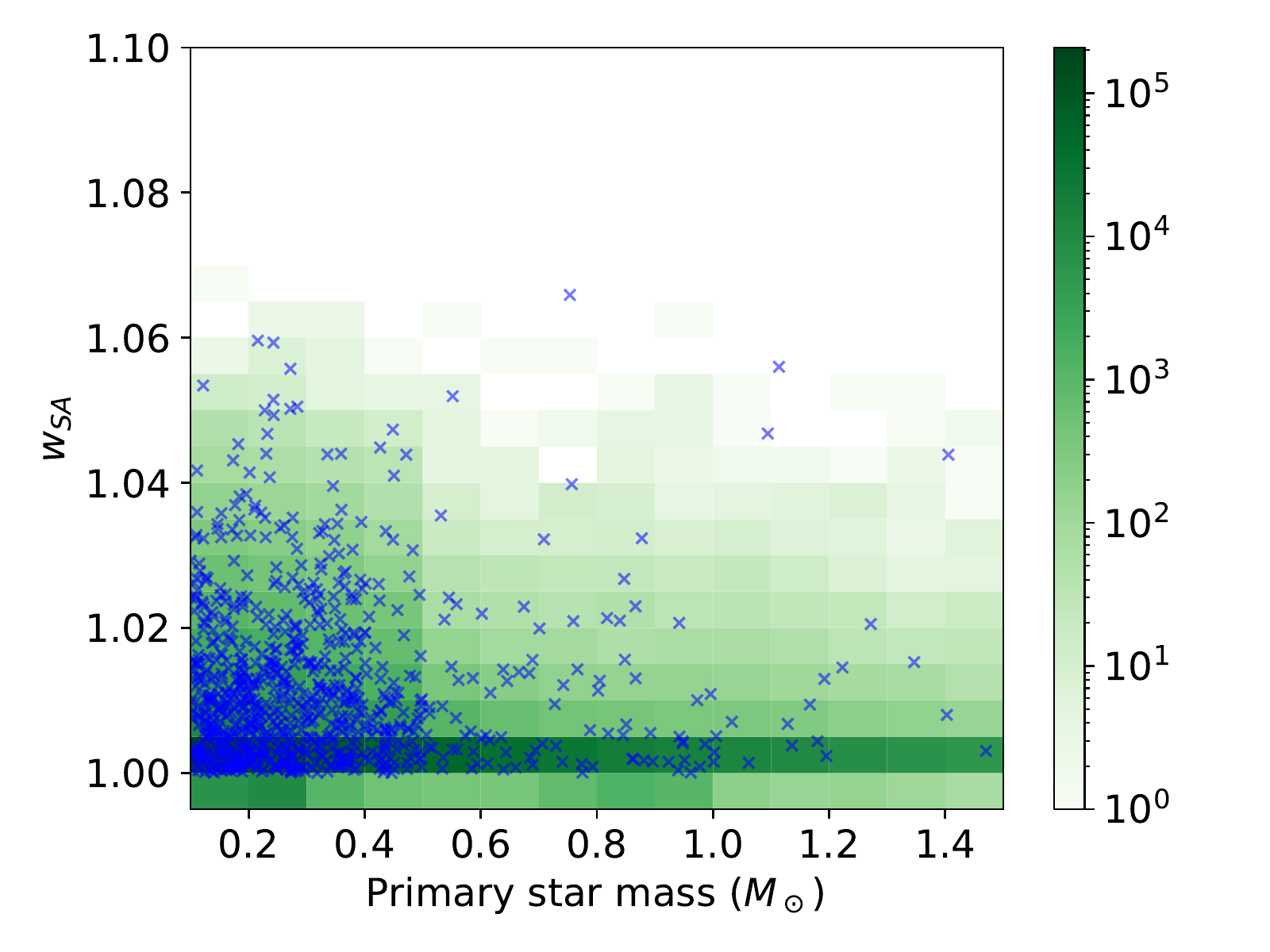}
\caption{Extension of the circumstellar HZs around the primary star in binary systems. Left panel: 2D histogram of $\Delta \ell_\text{SA}$ vs star mass for type-2 (blue) habitability condition (see Fig. \ref{HZtypes}). The gray-shaded area in the topmost part of the plot is empty due to the definition of the radiative HZ for a given star.
Right panel: 2D histogram of $w_\text{SA}$ vs star mass for type-1 (green) habitability condition, zoomed on the y-axis for better readability. Superimposed blue crosses represent type-2 cases with $w_\text{SA} \ge 1.0$.  Results obtained for Model A (Table \ref{tab:results}), from a simulation with $N_\circ=10^6$.
\label{fig_wSA}}
\end{figure}

To cast light on the extension of the HZs in binary systems, in Fig. \ref{fig_wSA} we plot the 2D histograms of the extension of the HZ around the primary. On the left panel, it is possible to see the absolute width $\Delta \ell_\text{SA}$ (in AU) for type-2 habitability conditions. It underlines the great range of widths of circumstellar HZs in binary systems, which spans from a negligible fraction of AU up to a few AU in the considered range of  parameters. This spread is expected due the fact that the boundary of dynamical instability can interrupt the radiative HZ at any location, according to all possible combinations of binary system parameters. There is a visible left-right density gradient induced by the primary mass distribution, which produces more small mass stars. An abrupt change at $m_\ell = 0.5$\,M$_\sun$ is also visible and it is caused by the different distribution of binary periods below and above this threshold. On the right panel, we show the normalized width $w_\text{SA}$ of type-1 HZs, with type-2 systems that sport a $w_\text{SA} \ge 1.0$ superimposed as blue crosses. As it is possible to see, the widths of most type-1 HZ are essentially equal to those around single stars (the string of dark green bins in the lower part of the panel) and depends on the fact that secondary stars usually play a minor role. On the other hand, there is a small number of systems with slightly enhanced HZ, in the order of some percent points. Also a small subset ($<0.1\%$) of type-2 cases is interested by this effect.

\begin{figure}[htbp!]
\plottwo{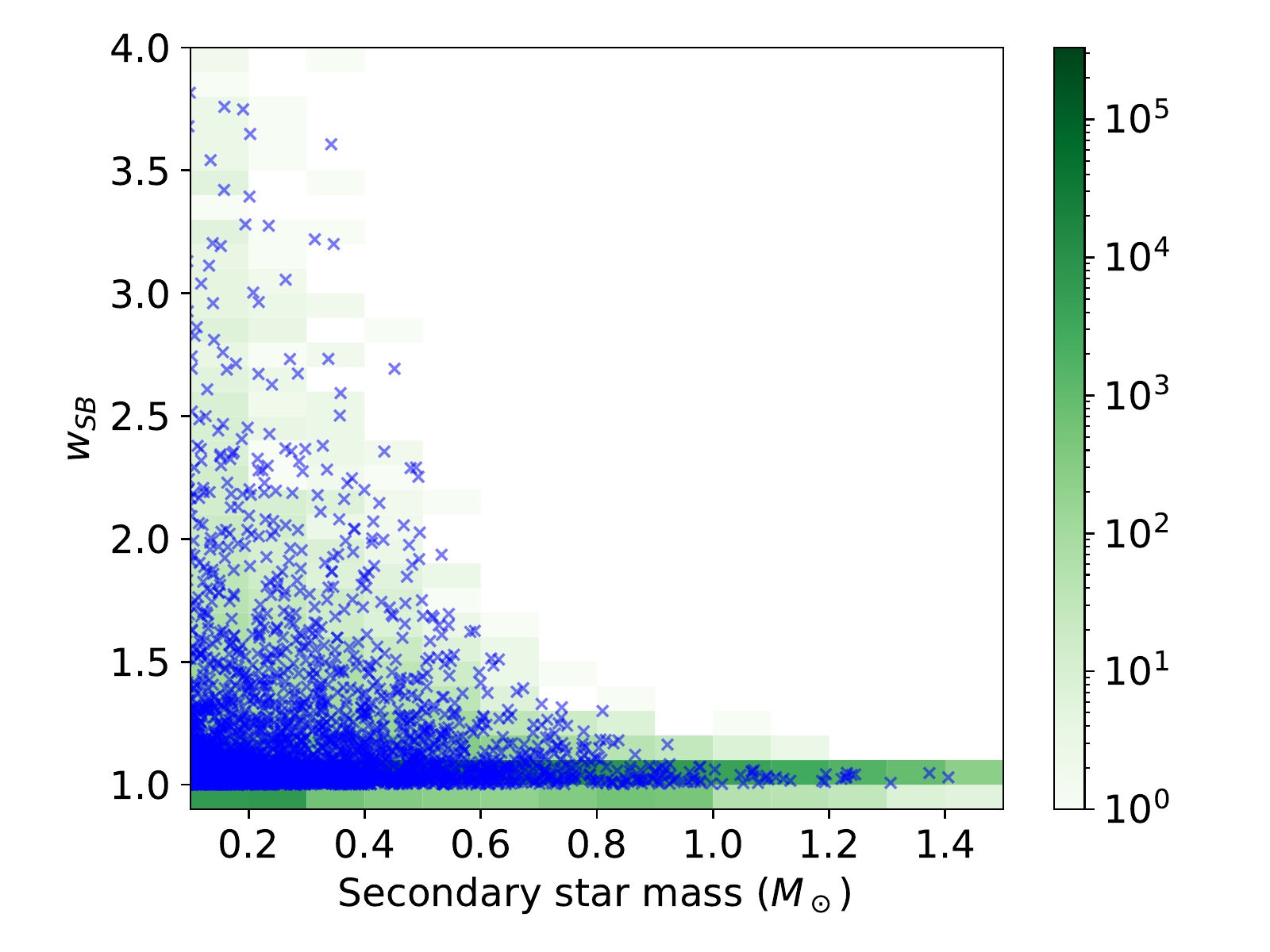}{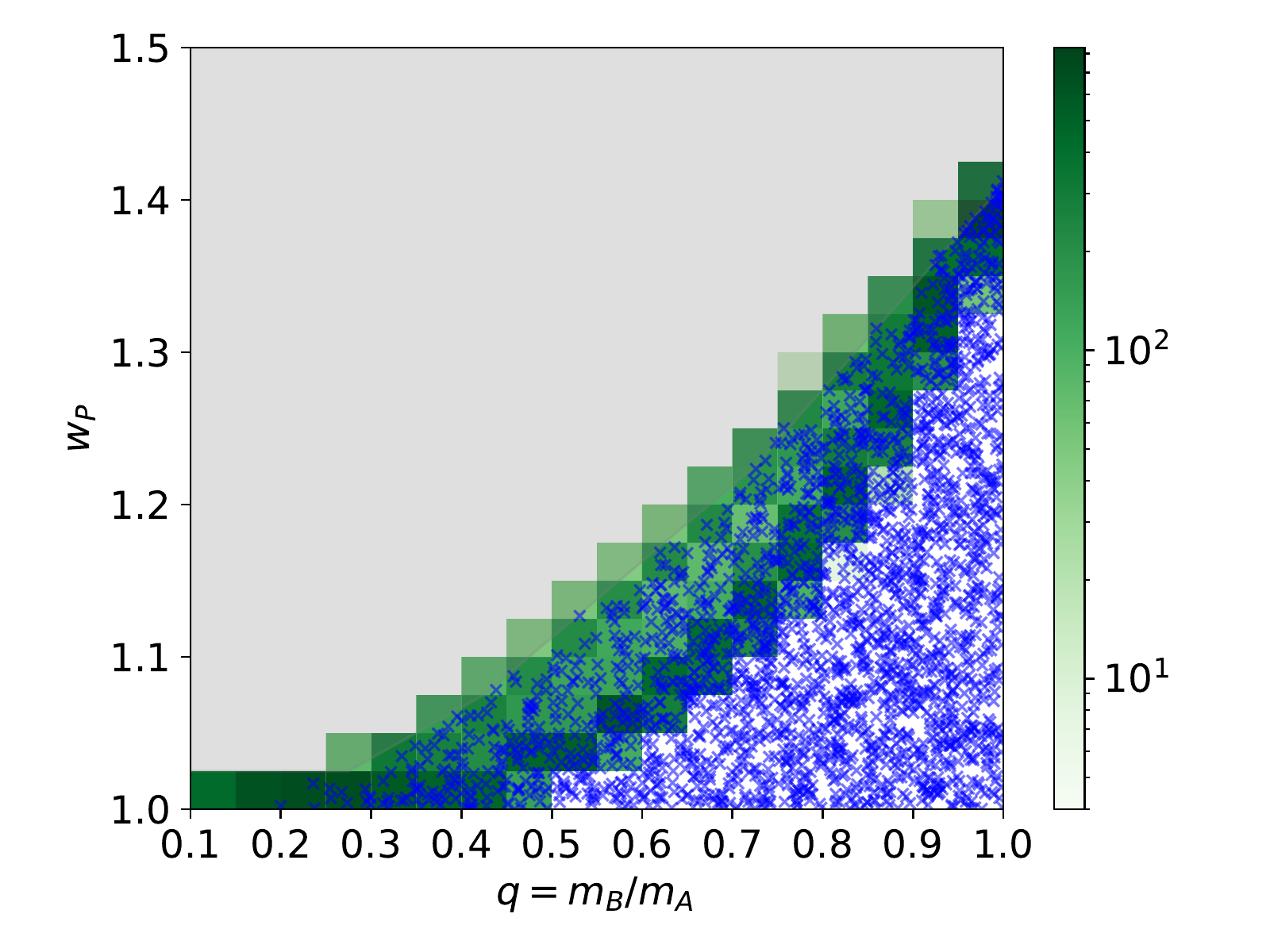}
\caption{Normalized width of HZs in binary systems. Left panel: 2D histogram of the normalized width of the circumstellar HZ around the secondary, $w_\text{SB}$,  versus mass of the secondary, $m_\text{B}$. Right panel: 2D histogram of the normalized width of the circumbinary HZ, $w_\text{P}$, versus mass ratio $q=m_\text{B}/m_\text{A}$. Green shades: type-1 conditions of habitability. Blue crosses: type-2 conditions of habitability with $w_\text{SB}$ or $w_\text{P}$ $\ge 1.0$. The gray-shaded area is inaccessible due to the definition of radiative HZ. Same sample as in Fig. \ref{fig_wSA}. 
\label{fig_wSB}}
\end{figure}

Concerning the width of the HZ around the secondary star, in the left panel of Fig.~\ref{fig_wSB} we plot $w_\text{SB}=\Delta \ell_\text{SB}/\Delta \ell_\circ (m_B)$ versus $m_B$. As in the primary star case analyzed above, most of the secondary stars sport HZs with comparable extensions as those around single stars of the same mass. Also in this case there are several instances of HZs wider than in the case of a single star, both for type-1 and type-2 habitability conditions. However, around secondary stars the enhancement effect is far more pronounced, especially for low mass stars with late type companions, where $w_\text{SB}$ can be as large as 7.4 and 5.8 for type-2 and type-1 HZs, respectively. An inspection of the systems where this result is found shows that this effect is due to the strong radiative flux of the primary which, in some cases, shifts outwards the insolation outer edge of the secondary. This effect explains why the circumstellar habitability around the secondary tends to be slightly larger than that around the primary, as we have seen when comparing Figs. \ref{fig_SA_ab_dist_hab} and \ref{fig_SB_ab_dist_hab}.

Concerning the width of circumbinary HZs, in the right panel of Fig.~\ref{fig_wSB} we plot the normalized width, $w_\text{P}=\Delta \ell_\text{P}/\Delta \ell_\circ (m_A)$, versus mass ratio $q=m_\text{B}/m_\text{A}$. One can see that also the circumbinary HZ can be wider than in the case of a single star, especially when the two stars have similar masses. This may happen not only for type-1 habitability conditions, but even when the radiative HZ is interrupted by the edge of dynamical instability (blue crosses). This effect becomes stronger with increasing mass ratio because the increasing contribution of radiative flux of the companion shifts the outer edge of the radiative HZ beyond the boundary of dynamical stability. The fraction of systems where this effect takes place is very small in the Monte Carlo sample. However, the effect leads to a significant extension of the circumbinary HZ in binary systems with stars of similar mass.
  
\subsection{Constraints from planetary formation}
\label{sectPlFormation}

\begin{deluxetable}{cccccccc}[b!]
\tablecaption{Statistical properties of habitability of the sub-samples obtained by considering: (i) limiting values in the binary semi-major axis $a_b$ for circumstellar and circumbinary planetary formation (columns 2 to 5) and (ii) different multiplicity fractions for M-type and FGK-type stars (columns 6 to 8). The properties of the whole sample for each model are shown in Table \ref{tab:results}.}
\tablecolumns{8}
\tablehead{ \colhead{Model}  & \colhead{$f_\text{hab,SA}^{>10\,\text{AU}}$} & \colhead{$f_\text{hab,SB}^{>10\,\text{AU}}$} & \colhead{$f_\text{hab,P}^{<1.0\,\text{AU}}$} & \colhead{$f_\text{hab,P}^{<0.3\,\text{AU}}$} & \colhead{$f_\text{hab,SA}^{\text{half M}}$} & \colhead{$f_\text{hab,SB}^{\text{half M}}$} & \colhead{$f_\text{hab,P}^{\text{half M}}$} }
\startdata
A & 0.998 & 0.999 & 0.171 & 0.368 & 0.831 & 0.860 & 0.043\\
B & 0.999 & 1.000 & 0.159 & 0.354 & 0.836 & 0.864 & 0.040\\
C & 0.997 & 1.000 & 0.200 & 0.418 & 0.822 & 0.865 & 0.047\\
D & 0.999 & 1.000 & 0.183 & 0.399 & 0.829 & 0.871 & 0.044\\
E & 0.997 & 0.999 & 0.185 & 0.389 & 0.819 & 0.851 & 0.047\\
F & 0.999 & 1.000 & 0.172 & 0.376 & 0.824 & 0.855 & 0.044\\
G & 0.997 & 1.000 & 0.214 & 0.440 & 0.809 & 0.855 & 0.052\\
H & 0.999 & 1.000 & 0.198 & 0.425 & 0.816 & 0.860 & 0.048\\
I & 0.994 & 0.998 & 0.210 & 0.415 & 0.779 & 0.829 & 0.055\\
J & 0.996 & 0.999 & 0.196 & 0.401 & 0.785 & 0.834 & 0.052\\
K & 0.992 & 0.998 & 0.226 & 0.440 & 0.765 & 0.819 & 0.060\\
L & 0.995 & 0.999 & 0.211 & 0.423 & 0.772 & 0.824 & 0.056\\
\enddata
\tablenotetext{}{}
\label{results_sub}
\end{deluxetable}

In Table \ref{tab:results} and  subsequent figures, we have analyzed the statistical properties of a general population of binaries under various hypotheses, regardless of the effective ability of a given binary configuration to give rise to a planetary system. While the details about planetary formation processes in binary systems are still under debate and fall out of the scope of the present work, we cannot ignore the influence that they may have on the statistics of habitable binaries. To deal with this issue, we extracted various sub-samples from each run, imposing limits on the binary semi-major axis, which seems to be the main factor in determining the limits of planetary formation in binaries. The results of this analysis, shown in columns 2, 3, 4 and 5 of Table \ref{results_sub}, are discussed below.

\subsubsection{Planetary formation in circumstellar orbits}
\label{sectPlFormation_S}

The ability to form planets from a circumstellar protoplanetary disk perturbed by a companion star is a topic investigated both numerically \citep{Quintana07}, analytically \citep{Rafikov15a,Rafikov15b,Silsbee15} and observationally \citep{Kraus16}. Theoretical models indicate that the formation and growth of Earth-mass planets in the radiative HZ of a G-type primary star is generally allowed when $a_b \gtrsim 10$\,AU.
However, they also allow for the formation of close-in planets when $a_b = 5$ AU \citep{Quintana07} or the formation in the radiative HZ when certain specific conditions on orbital parameters are met \citep{Rafikov15b}. From the observational point of view, \citet{Kraus16} found a dearth of planet candidates around binaries with $a_b \le 50$ AU and no candidates below $a_b \sim 10$ AU. Finally, \citep{Barnes20} identified a circumstellar super-Earth around HD 42936, whose binary separation is $1.22$ AU. Here we consider this latter  system as an outlier and we adopt a tentative lower limit of $a_b > 10$ AU for circumstellar planetary formation. We extract a sub-sample with this cutoff for each of our models and we analyse the results.
As it is possible to see in Table \ref{results_sub} (columns $2$ and $3$), the adoption of this cutoff increases $f_\text{hab,SA}$ and $f_\text{hab,SB}$ to near 100\% for every model. Therefore, the constraints of dynamical stability of already formed planets are less stringent than those of circumstellar planetary formation. These latter constraints, albeit still uncertain at the present time,  suggest that every binary configuration where circumstellar planets can be formed allows habitable circumstellar planets to exist.

\subsubsection{Planetary formation in circumbinary orbits}
\label{sectPlFormation_P}

Planetary formation in circumbinary protoplanetary disks also has some issues. \citet{Quintana06} investigated this topic via numerical simulations finding that, for binary apastrons $\lesssim 0.2$\,AU, planetary formation proceeds in the same way as around single stars, while for apastrons $>0.3$\,AU the formation of Earth-like planets at 1 AU is less likely. On the observational side, \citet{Trilling07} found that a large fraction of young binaries with separations $< 3$\,AU have circumbinary disks, with at least some instances of dynamically stable ones for $a_b > 1$ AU. \citet{Alves19} demonstrated the undergoing dissipation of the circumbinary disk around Barnard 59, whose separation is $\sim 20$\,AU. Finally, the widest circumbinary planet-hosting binary known up to date has $a_b=0.23$\,AU \citep{Welsh15}. Clearly, it is difficult to establish a hard upper limit in $a_b$ for circumbinary planetary formation at the present time. Given that the fraction of circumbinary planet-hosting systems is strongly dependent on the adopted cutoff, we extracted two sub-samples, one with $a_b < 1.0$\,AU and another with $a_b < 0.3$\,AU. In these two sub-samples $f_\text{hab,P}$ increases considerably, from $3-5\%$ of the complete sample, up to $16-23\%$ or $35-44\%$ (columns 4 and 5 in Table \ref{results_sub}, respectively).

In any case, even the most selective sub-sample still leaves the potentially habitable, circumbinary planet-hosting systems in minority. This points to an intrinsic paucity of circumbinary habitable planets, regardless of the specific hypotheses about the $q$, $a_b$ and $e_b$ distributions.

Alongside the upper limit in the binary star separation, some authors pointed also to the possible existence of a lower limit for this variable. \citet{Martin15} and \citet{Martin18} underlined a paucity of exoplanet detections in very close binary stars with periods of less than 5 days. They suggest that the presence of a third, undetected faint companion in some of the members of the sample could have skewed their conclusion. On the other hand, \citet{Fleming18} explained this dearth through an interplay of successive angular momentum transfers (between the two stars and the newly formed planets) and magnetic braking. Circumbinary planets would then exist only for systems with $P_b$ over 5 days. If we adopt this lower limit for binary star separation we find a reduction of $20-25 \%$ of $f_{\text{hab,P}}$ across all our models. This fact, once again, reinforces our conclusion that circumbinary habitable planets are rare.

\subsection{Effects of the short-period binary excess}

From a reanalysis of the Kepler Eclipsing Binary Catalog, \citet{Kirk16} found an excess of very-short period (in the $0.3-1$ day interval in $P_b$) binary systems  that is not captured by the log-normal distribution that we have adopted (see Sect.~\ref{sectPeriods}). This excess amounts to $\sim 0.5\%$ of the total population. Given that most of the binaries with $P_b < 10$ days host type-1 circumbinary HZs, this in turn increases the fraction $f_{hab,P}$.  
This does not affect our conclusions since the increase is quite modest across our models, affecting only the last decimal digit of the values in Table \ref{tab:results}. It is important to recall that in such close systems, a variety of complex phenomena set in, like e.g.~Roche lobe overflows, which can substantially change the radiative properties of the stars involved. Our code is not equipped for  studying these particular environments.

\subsection{Binary occurrence rate versus spectral type of the primary}
\label{sectOccurrence}

In the methodology that we designed to generate samples of binaries we extract the mass of the primary star from the IMF or the PDMF distributions described in Section \ref{sectPrimaryMass}. In doing so, we implicitly assume that the occurrence rate of binaries is the same across different spectral types (FGKM) of the primary. This is the simplest approximation given the uncertainties surrounding this topic. However, there are clues that this could not be the case. The synthesis of results analysed by \citet{Raghavan10} indicate that the occurrence rate of binaries increases with mass, with multiplicity fraction of $42-48\%$ in the F6-K3 range, compared to $11-42\%$ for M-dwarfs \citep{Fischer92,Reid97}. \citet{Moe17} observed a similar trend in the O-G range. Considering the uncertain distinction between the binary and the higher-order multiplicity fractions, it is difficult to quantify these trends.
To assess the impact of a possible dearth of binary systems with red dwarf primaries, we extracted a subset from each of the main samples, removing half of the systems with M-type primaries. As a result, we find that the circumstellar habitability fractions $f_\text{hab,SA}$ and $f_\text{hab,SB}$ of the subsets (Table \ref{results_sub}, columns 6 and 7)  are very similar to those of the whole samples (Table \ref{tab:results}). The circumbinary habitability fractions $f_\text{hab,P}$ of the subsets  tend to increase, but remain very low, in the order of $4- 6\%$ (Table \ref{results_sub}, column 8).

Finally, we combined the constraints on $a_b$ for planetary formation with the reduction in the binary occurrence rate for M-type primaries. In this case, we do not report the results in the table. For circumstellar habitability, both $f_\text{hab,SA}$ and $f_\text{hab,SB}$ are virtually identical to those reported in columns $2$ and $3$ of Table \ref{results_sub} in all models. On the oher hand, the fractions of circumbinary habitability tend to increase slightly for both of the adopted limits for planetary formation. Even in this scenario, $f_\text{hab,P}$ is generally $ \lesssim 50\%$, confirming the paucity of circumbinary configurations suitable for life. The higher results in this regard are achieved by models G and K, that reach a $f_\text{hab,P}$ value equal to 54\% and 55\%, respectively. This analysis suggests that our results are weakly affected by the uncertainties of the stellar orbital parameters and occurrence rates in binary systems.

\subsection{Model dependence of the results} 
\label{sectModDep}

The fact that the properties of circumbinary and circumstellar habitability in binary systems are similar for most models lends support to the robustness of the main conclusions that we have summarized above.
To cast light on the model dependence of our procedure, we now discuss some minor, but systematic differences that are seen when comparing the results obtained from different models. 

\subsubsection{Distribution functions}
\label{sectDistributionFunctions} 

When passing from the IMF to the PDMF, keeping fixed the other prescriptions, our results tend to give slightly higher habitability to circumstellar regions and slightly lower habitability to circumbinary regions. This can be seen in Table \ref{tab:results}, where a change from IMF to PDMF yields an increase of habitable fractions $f_\text{hab,SA}$ and $f_\text{hab,SB}$ and a small but systematic rise of $\overline{w}_\text{SA}$ and  $\overline{w}_\text{SB}$. Conversely, for circumbinary regions we see a small decrease of $\overline{w}_\text{P}$ and a small decrease of the habitable fraction  $f_\text{hab,P}$. These changes can be interpreted in terms of: (1) the lower contribution of high-mass stars in the PDMF and (2) the strong dependence of stellar luminosity on stellar mass on the main sequence. The lower fraction of high-luminosity stars in the PDMF implies that, on average, the radiative HZ will get closer to the stars. As a result, the radiative HZ will get closer to the circumstellar region of stability, increasing the habitability of S-type orbits, and closer to the circumbinary region of instability, decreasing the habitability of P-type orbits.

As for the distribution of mass ratios, $\xi(q)$, it is possible to see a minor but systematic increase in the circumstellar habitable fractions $f_\text{hab,SA}$ and $f_\text{hab,SB}$ when we use the top-heavy DK13 power-law instead of the log-normal, low-q centered function of DM91. This is unexpected, given that systems with similar-sized stars should have narrower dynamical stability zones around the primary and further out inner HZ edges around the secondary, other things being equal.
 
As far as the distribution function of eccentricities of stellar orbits, $\xi(e_b)$, is concerned, using the results of \citet{Moe17} yields an higher value for $f_\text{hab,P}$ and a smaller value for both $f_\text{hab,SA}$ and $f_\text{hab,SB}$ than using the \citet{StepinskiBlack01} prescription. In this case, it is not easy to disentangle the effect of changing $\xi(e_b)$ from the effect of changing $\xi(q)$. Even if MDS17 $\xi(q)$ gives rise to a final $q$ distribution that is largely similar to the DM91 one, the former introduces a correlation with $P_b$ that is absent in the latter. Nonetheless, a detailed analysis of the sample shows that the increase in the circumbinary habitable fraction is caused by the correlation between binary eccentricity and period that favors the generation of low eccentricity systems when the period is small. This in turn allows for smaller possible planetary semi-major axes. On the other hand, for larger binary periods, larger eccentricities are favored, and this cause the small reduction of the circumstellar habitable fractions around both stars.

The above considerations on the adopted distribution functions are also valid for the subsamples defined in \ref{sectPlFormation} and \ref{sectOccurrence}, as it is possible to see in Table \ref{results_sub}. In particular, for S-type orbits, most of the variations are swamped by the fact that both $f_\text{hab,SA}$ and $f_\text{hab,SB}$ have values near unity, while for $f_\text{hab,P}$ they are amplified and more clearly visible, especially when the planetary formation limits are considered.

\begin{deluxetable}{ccccccccc}[b!]
\tablecaption{Comparison of the indices of habitability obtained by adopting the recipes for $a_\text{crit}$ from \citet{Quarles18,Quarles20} and the original expressions from \citet{HW99}.}
\tablecolumns{9}
\tablehead{ \colhead{Model$^a$} & \colhead{$e_p$} & \colhead{$f_\text{hab,SA}$} & \colhead{$\overline{w}_\text{SA}$} & \colhead{$f_\text{hab,SB}$} & \colhead{$\overline{w}_\text{SB}$} & \colhead{$f_\text{hab,P}$} & \colhead{$\overline{w}_\text{P}$} & \colhead{Reference for $a_\text{crit}$} }
\startdata
 A & 0.0 & 0.843 & 0.955 & 0.869 & 0.970 & 0.035 & 0.930 & \citet{Quarles18,Quarles20} \\ 
 A & 0.0 & 0.857 & 0.960 & 0.885 & 0.974 & 0.031 & 0.939 & \citet{HW99} \\
 B & 0.0 & 0.848 & 0.955 & 0.874 & 0.969 & 0.034 & 0.929 & \citet{Quarles18,Quarles20} \\ 
 B & 0.0 & 0.863 & 0.960 & 0.889 & 0.973 & 0.029 & 0.932 &  \citet{HW99}  \\  
 C & 0.0 & 0.835 & 0.955 & 0.874 & 0.979 & 0.039 & 0.885 & \citet{Quarles18,Quarles20}\\  
 C & 0.0 & 0.848 & 0.959 & 0.888 & 0.983 & 0.034 & 0.895 &  \citet{HW99} \\
 D & 0.0 & 0.839 & 0.955 & 0.877 & 0.977 & 0.036 & 0.882 & \citet{Quarles18,Quarles20} \\ 
 D & 0.0 & 0.854 & 0.959 & 0.892 & 0.981 & 0.031 & 0.886 &  \citet{HW99} \\
\enddata 
\tablenotetext{a}{Distribution functions adopted in each model are shown in Table \ref{tab:results}. $N_\circ = 10^5$.}
\label{comparisonHW}
\end{deluxetable}

\subsubsection{Boundaries of dynamical stability}

The exact location of the boundaries of dynamical stability $a_\text{crit}$ play an essential role in our calculations. The determination of the regions of dynamical instability depends on the accuracy of the N-body simulations performed by different authors. Our results are based on the analytical expressions \eqref{acritP} and \eqref{acritS} derived by \citet{Quarles18,Quarles20} from a large database of simulations, each simulation covering $10^5$ orbital periods of the binary system. To assess the impact of the uncertainties related to the definition of the regions of stability, we repeated our calculations adopting the original analytical expressions derived by \citet{HW99} from a more limited set of N-body simulations covering $10^4$ orbital periods. The indices of habitability obtained from the two different recipes of $a_\text{crit}$ are compared in Table \ref{comparisonHW}. One can see that the results match well, the differences being consistent with the statistical uncertainties. This is true also for the subsamples described in Sections \ref{sectPlFormation} and \ref{sectOccurrence} (not reported in Table \ref{comparisonHW}).

\subsubsection{Insolation boundaries of the radiative HZ}

The results presented in Table \ref{tab:results} have been derived using the climatological limits RG and MG. Adopting the empirical limits RV and EM we find a slightly higher habitability, but  the results do not change significantly. In circumbinary regions, where the habitability is very sensitive to the location of the outer edge of insolation, the differences are particularly small because the MG edge is quite close to the EM edge (Table\,\ref{coeffs}). Indeed, we find that the fraction of habitable circumbinary regions increases typically by $\sim 0.1-0.2\%$ adopting the empirical limit EM. The differences are more marked in the circumstellar regions, which  are particularly sensitive to the location of the inner edge, because the RV is significantly closer to the star than the RG edge (Table\,\ref{coeffs}). We find that the fraction of habitable circumstellar regions increases by $\sim 2-3\%$ adopting the empirical limit RV. These differences do not affect  the conclusions of the present work. In the rest of the paper we use the results obtained from the climatological limits RG and MG as a reference for our discussion. Similarly modest variations are found in the $a_b$-limited and the binary M-dwarf halved samples described in Sections \ref{sectPlFormation} and \ref{sectOccurrence}.

\subsection{Eccentricity of planetary orbits}
\label{sectTrendPlanEcc}

The eccentricity of planetary orbits affects the habitability of binary stellar systems both in terms of insolation and in terms of dynamical stability. In terms of insolation, we know that the mean stellar flux received by a planet moving in an elliptic orbit with eccentricity $e_p>0$ is enhanced by a factor $(1-e_p^2)^{-1/2}$ compared to the case of circular orbits \citep[see, e.g.,][]{Williams02,Vladilo13}.
As we have already emphasized, in binary systems an enhancement of insolation has opposite effects of habitability in circumstellar and circumbinary regions. In circumstellar regions the larger insolation yields a larger frequency of type 3 habitability conditions, whereas in circumbinary regions a larger frequency of type 1 conditions. Therefore we expect a decrease of circumstellar habitability and an increase of circumbinary habitability with increasing $e_p$. Even though planetary eccentricity may be forced to low values both by dynamical stability considerations and by momentum exchange between the stars and the planet, in order to quantify the potential impact on insolation, we repeated our simulations applying a correction term $(1-e_p^2)^{-1/2}$ to the stellar luminosities. For P-type orbits, we applied this correction to both stars since stable planetary orbits are far from both stars. For S-type orbits, we  applied the insolation enhancement only to the host star. Results for model A with increasing value of $e_p$ are shown in Table \ref{planetaryEccentricity}. With increasing planetary eccentricity one can see, as expected, a decrease of $f_\text{hab,SA}$ and $f_\text{hab,SB}$ and an increase of $f_\text{hab,P}$. Correspondingly, we see variations of the normalized width of the HZs, with a decrease of $\overline{w}_\text{SA}$ and $\overline{w}_\text{SB}$ and an increase of $\overline{w}_\text{P}$. The same happens in the subsamples described in Sections \ref{sectPlFormation} and \ref{sectOccurrence} (not reported in Table \ref{planetaryEccentricity}).  These changes are modest, at least up to $e_p \simeq 0.6$. However, the dynamical stability of configurations with $e_p \gtrsim 0.6$ are critical in most of the cases. Planetary orbital eccentricity, if not damped out by momentum exchange, may hardly increase the habitability of circumbinary regions, in any case only up to the critical conditions of stability. 

\section{Discussion}\label{sec:discussion}

We now discuss the potential impact of our results on searches for habitable planets in binary systems. First we investigate which observable properties of binary systems can be used to optimize searches for the highest conditions of circumstellar or circumbinary habitability. Then we discuss potential applications to known exoplanets and for planning future searches of exoplanets. 

\subsection{Searching for binary systems with habitable zones}

\subsubsection{Trends with binary system properties}
\label{sectTrendBinaryOrbitalParameters}

\begin{figure}[htbp!]
\plottwo{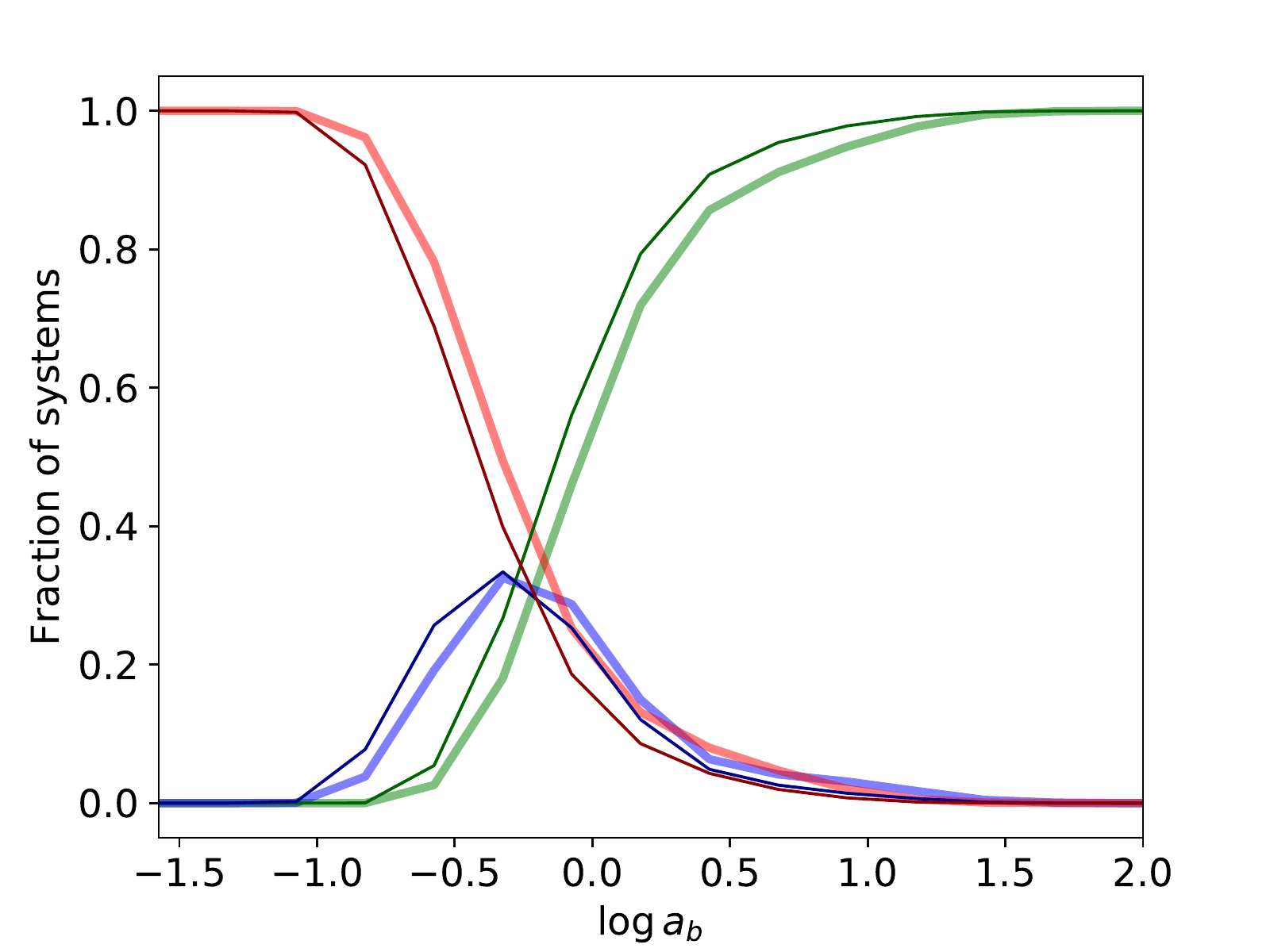}  {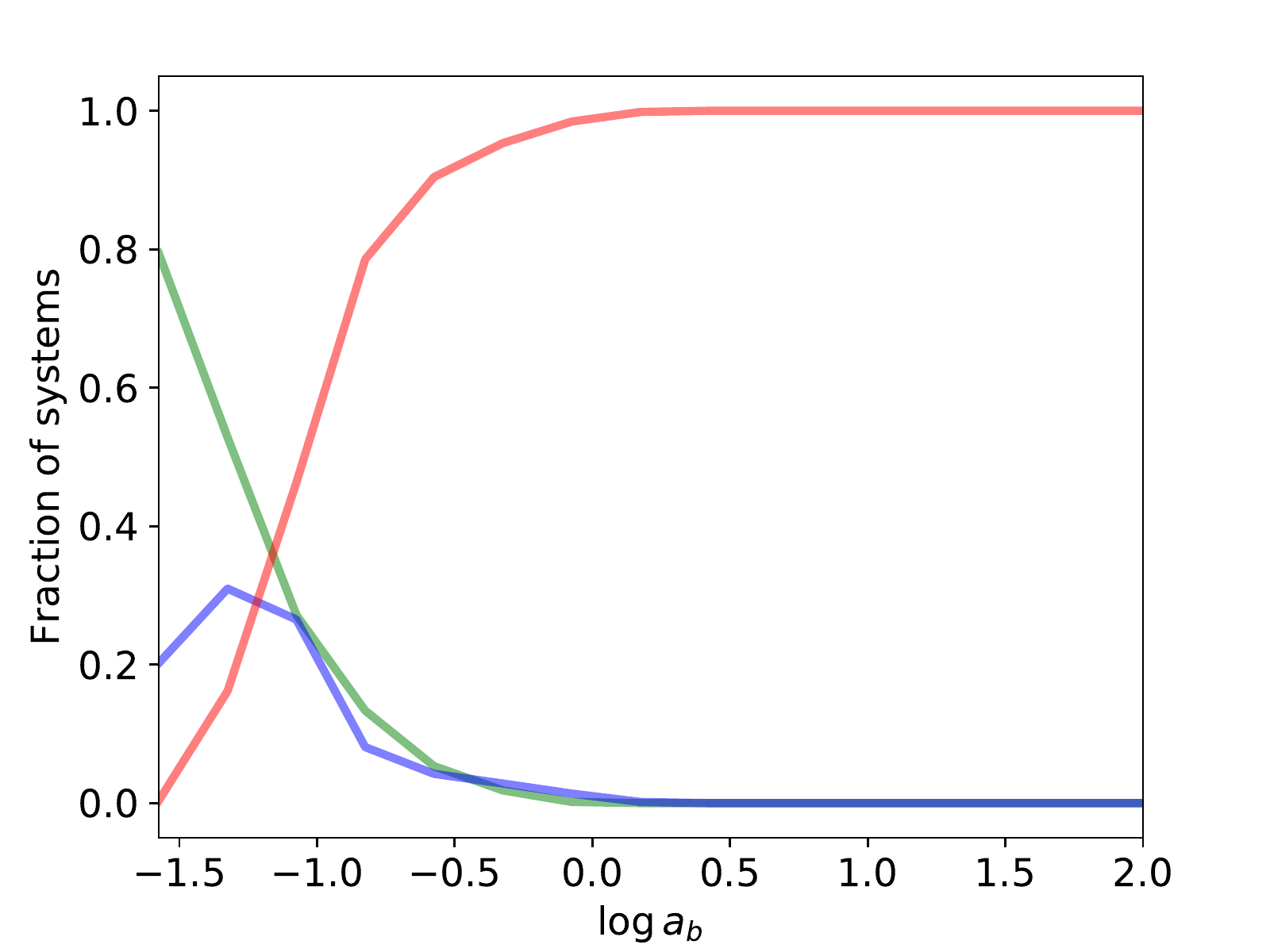} 
\caption{Fraction of systems with different conditions of habitability as a function of binary semimajor axis. In each bin of $\log a_b$ the number of systems with type-1 (green), type-2 (blue) and type-3 (red) conditions of  habitability is normalized to the total number of systems in the same bin. The plots are truncated to $\log a_b=2$ to better highlight the trends. Left panel: circumstellar regions around the primary (thick light lines) and the secondary star (thin dark lines). Right panel: circumbinary regions. Same Monte Carlo sample as in Fig. \ref{fig_SA_ab_dist_hab}.   
\label{fig_norm_hist_ab}}
\end{figure}

The stellar separation has a strong influence on the binary habitability  because the boundaries of dynamical stability scale  with the semi-major axis of the binary system, $a_b$, as shown in Eqs. (\ref{acritP}) and (\ref{acritS}). In circumstellar regions, when $a_b$ is sufficiently small the largest stable orbit will be so close to the star to have an insolation above the Runaway Greenhouse limit. In circumbinary regions, when $a_b$  is sufficiently small, stable orbits of P type will be so close to both stars that the planets will have sufficient insolation to escape the snowball state. The impact of these effects can be seen in Fig. \ref{fig_norm_hist_ab}, where we show the fraction of systems with type-1,-2 and -3 conditions of habitability plotted in constant bins of $\log a_b$. One can see that the fraction of circumprimary type-1 regions increases with $a_b$, from $\simeq 0$ at $a_b < 0.1$ AU up to $\simeq 1$ at $a_b > 10$ AU, with a sharp rise at $a_b \simeq 1$ AU. In addition, there is a small fraction of systems with type-2 conditions between $\simeq 0.1$ and $\simeq 3$ AU. If we include the constraints of planetary formation (Section \ref{sectPlFormation}), nearly any system that can form planets also has a dynamically stable radiative HZ. One can see that circumsecondary (thin dark lines) and circumprimary (thick light lines) habitability show similar values and trends. In the right panel we show the results for circumbinary regions, where the behaviour is opposite to that of circumstellar regions. Specifically, circumbinary habitability starts very high at $a_b < 0.1$ AU, but then declines sharply with increasing $a_b$. Since the limit for planetary formation in circumbinary orbits seems to lie in the $0.3-1.0$ AU range, this result shows that a sizeable number of systems able to form circumbinary planets cannot support a stable radiative HZ. This fact clearly differentiates P-type habitability from S-type habitability.

\begin{figure}[htbp!]
\plottwo{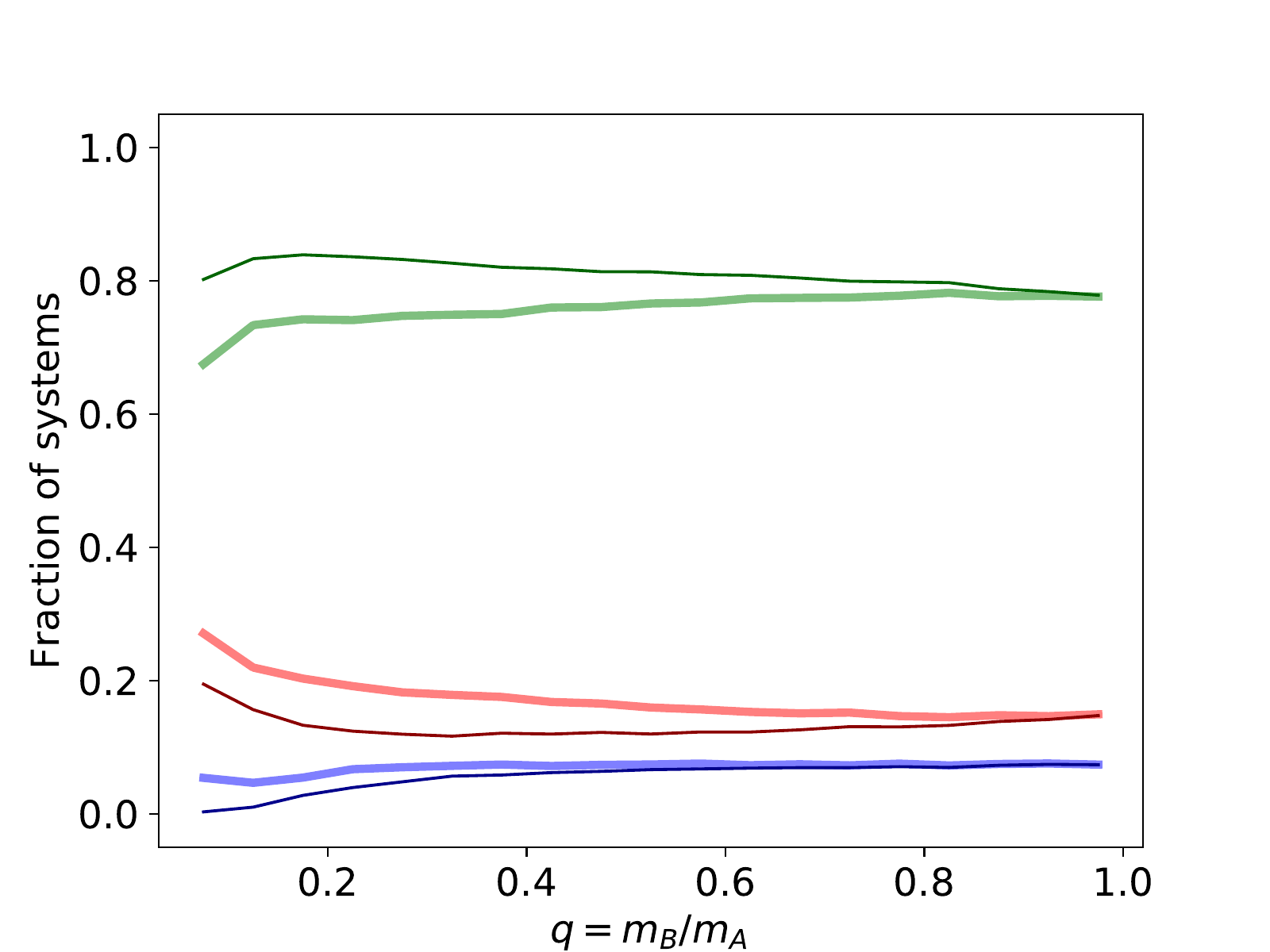}  {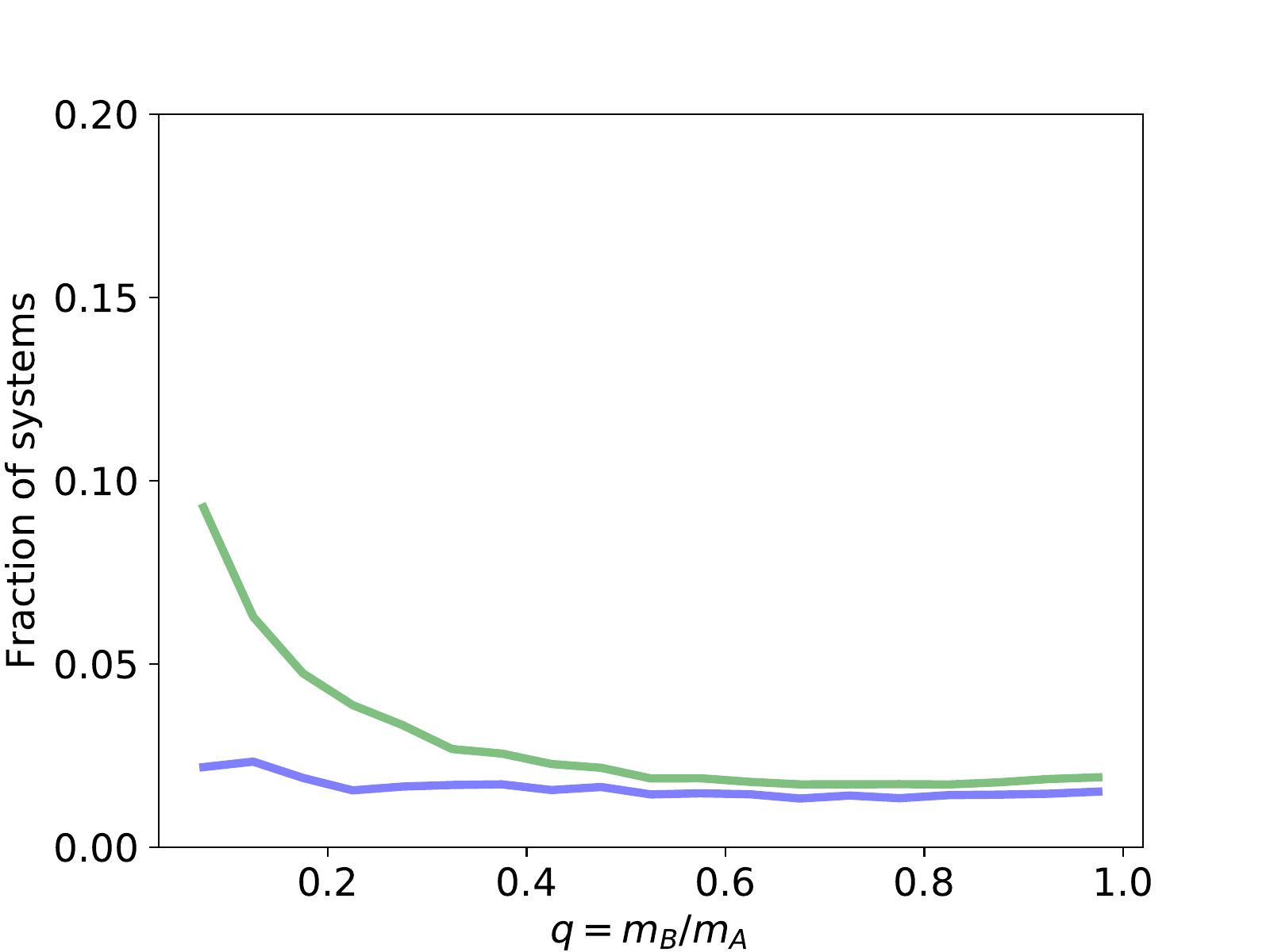} 
\caption{Fraction of systems with different conditions of habitability as a function of mass ratio, $q=m_\text{B}/m_\text{A}$. In each bin of $q$ the number of systems with type-1 (green), type-2 (blue) and type-3 (red) conditions of  habitability is normalized to the total number of systems in the same bin. Left panel: circumstellar regions around the primary (thick light lines) and  the secondary star (thin dark lines). Right panel: circumbinary regions (type-3 HZ are omitted to better highlight the trend for type-1 and 2 HZs). Same Monte Carlo sample as in Fig. \ref{fig_SA_ab_dist_hab}.   
\label{fig_norm_hist_mu}}
\end{figure}

In Fig. \ref{fig_norm_hist_mu}  we show the impact of the mass ratio, $q=m_\text{B}/m_\text{A}$, on the relative number of circumstellar and circumbinary HZs. In the left panel one can see that the fraction of cases with type-1 conditions around the primary shows a modest rise with increasing $q$, with a maximum value $\simeq 0.8$ when both stars have the same mass (thick light lines). For the circum-secondary regions the fractions of different types of habitability conditions is almost independent of $q$ (thin dark lines). Also the fractions of circumbinary HZs show no significant trend with $q$, apart from a very modest increase towards smaller mass ratios. This result, which concerns the number of systems in a given condition, should not be confused with the result concerning the extension of the circumbinary HZ which becomes larger with increasing $q$, as we have already seen in Fig. \ref{fig_wSB}. Therefore, the mass ratio is not a strong discriminant as far as the relative number of cases in different habitability conditions is concerned, but it does have an impact on the extension of the HZs. 

\begin{figure}[htbp!]
\plottwo{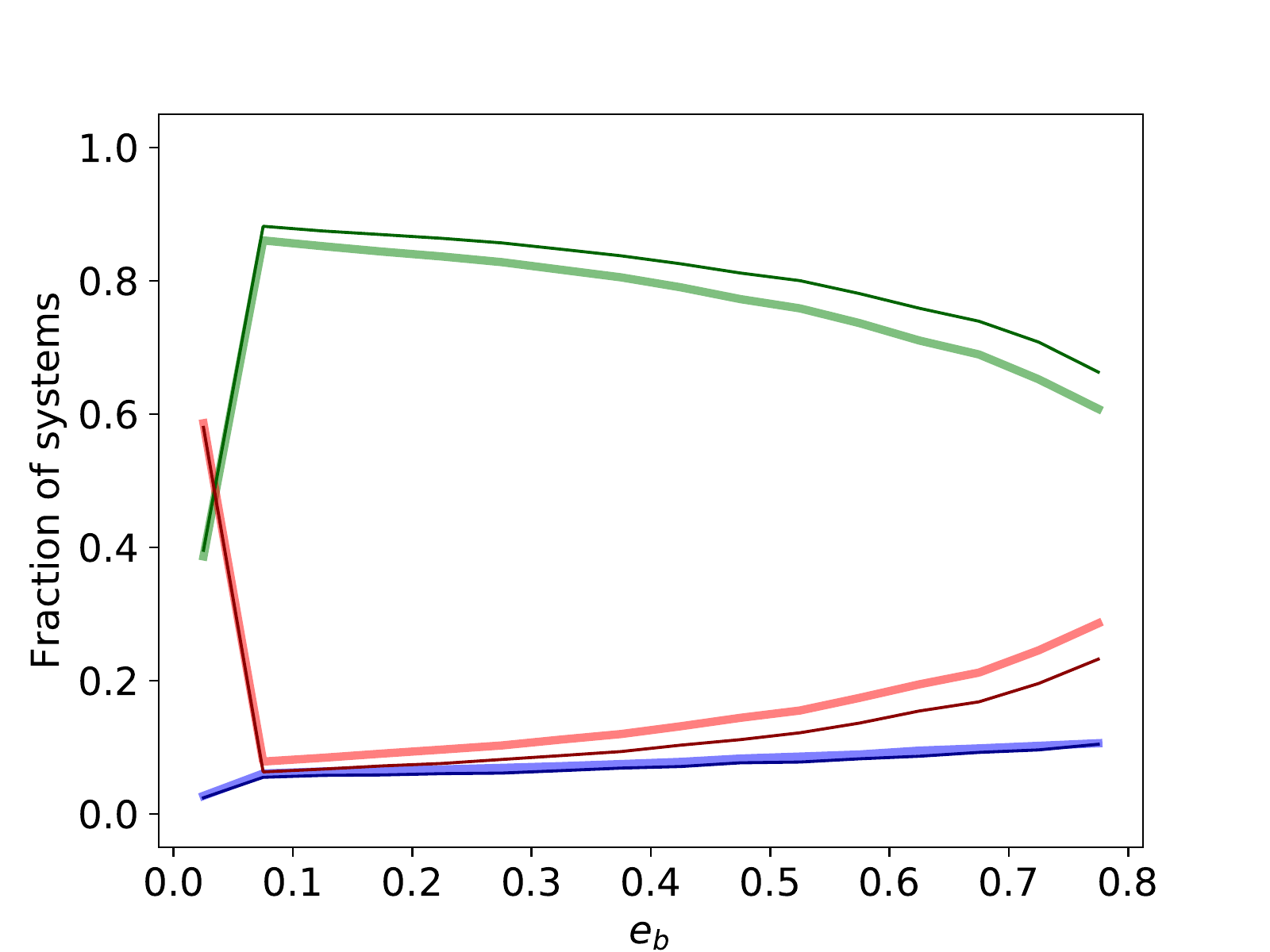}  {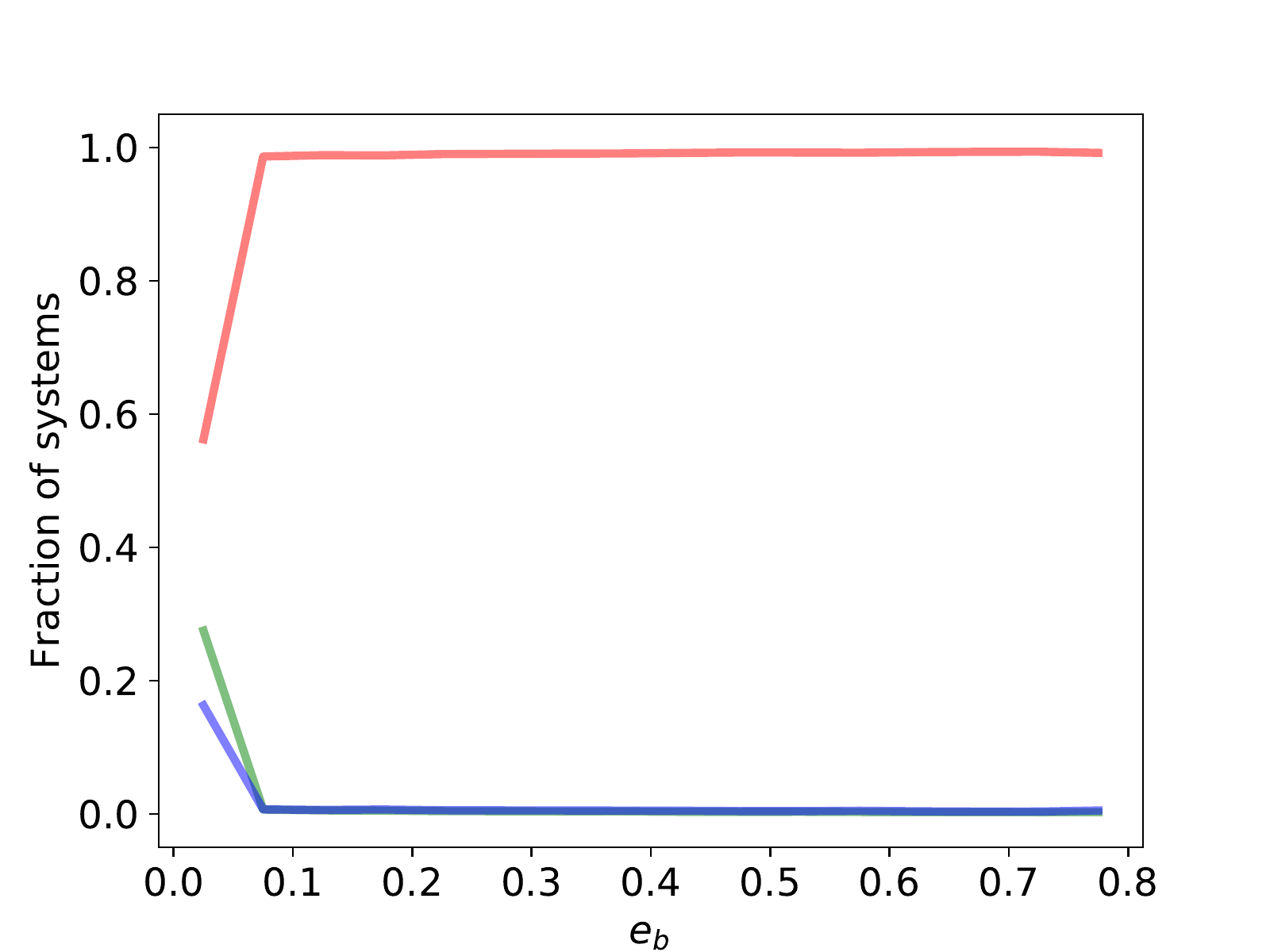} 
\caption{Fraction of systems with different conditions of habitability as a function of the binary orbital eccentricity. In each bin of $e_b$ the number of systems with type-1 (green), type-2 (blue) and type-3 (red) conditions of  habitability is normalized to the total number of systems in the same bin. Left panel:  circumstellar regions around the primary (thick light lines) and the secondary star (thin dark lines). Right panel: circumbinary regions. Same Monte Carlo sample as in Fig. \ref{fig_SA_ab_dist_hab}.   
\label{fig_norm_hist_eb}}
\end{figure}

In Fig. \ref{fig_norm_hist_eb}  we show the impact of the stellar orbital eccentricity, $e_b$, on the relative number of binary HZs. For circumstellar regions around the primary (left panel, thick light lines) one can see a sudden, initial rise of the fraction of HZs with increasing $e_b$, followed by a smooth decline. Similar results are found for regions around the secondary (left panel, thin dark lines). Also in this case we find an opposite behaviour in circumbinary regions (right panel), with a sudden decrease of the fraction of HZs when $e_b$ increases above $0.0$. The sudden changes above $0.0$ are due to the fact that the stellar orbits are assumed to be circularized ($e_b=0$) at short orbital periods. This means that at $e_b=0$ the semimajor axis is very small and so, as discussed above, circumstellar regions become less habitable, whereas circumbinary regions more habitable. As soon as $a_b$ gets larger, $e_b$ can assume any value, including larger values, and this creates a sudden change of habitability. This happens regardless of the specific choice of $\xi(e_b)$  and is magnified by the use of normalized bins. On top of this effect, there is a trend due to the fact that the boundaries of dynamical stability vary with $e_b$ according to Eqs. \eqref{acritP} and \eqref{acritS}. With increasing $e_b$ the circumstellar edge $a_\text{crit,S}$ gets closer to the host star, whereas the circumbinary edge $a_\text{crit,P}$ is shifted outwards. As a result both the circumstellar and the circumbinary regions become less habitable, but this effect is modest.

\subsubsection{Trends with stellar spectral type}  
\label{sectTrendStellarProperties}

\begin{figure}[ht!]
\plottwo{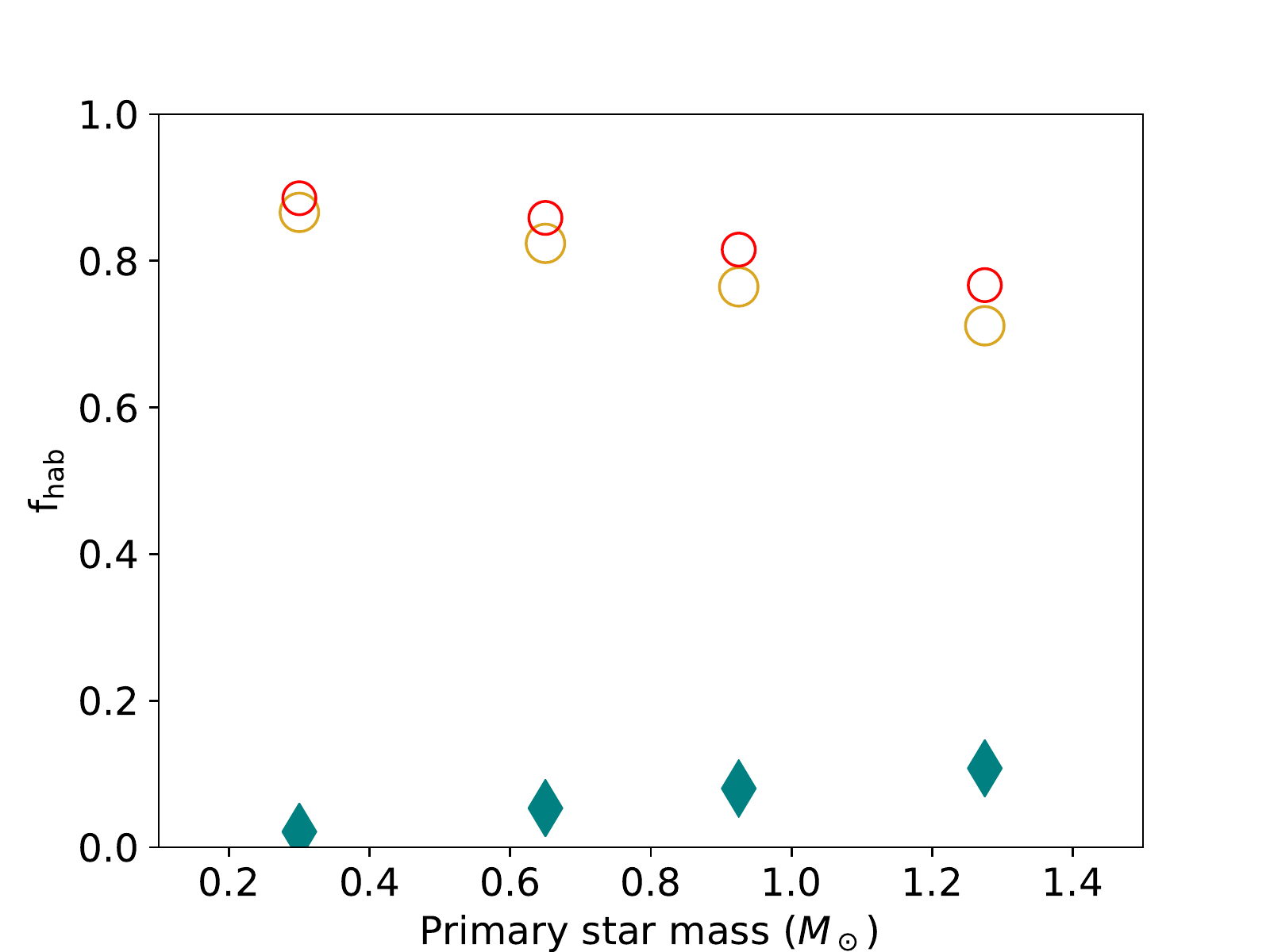}{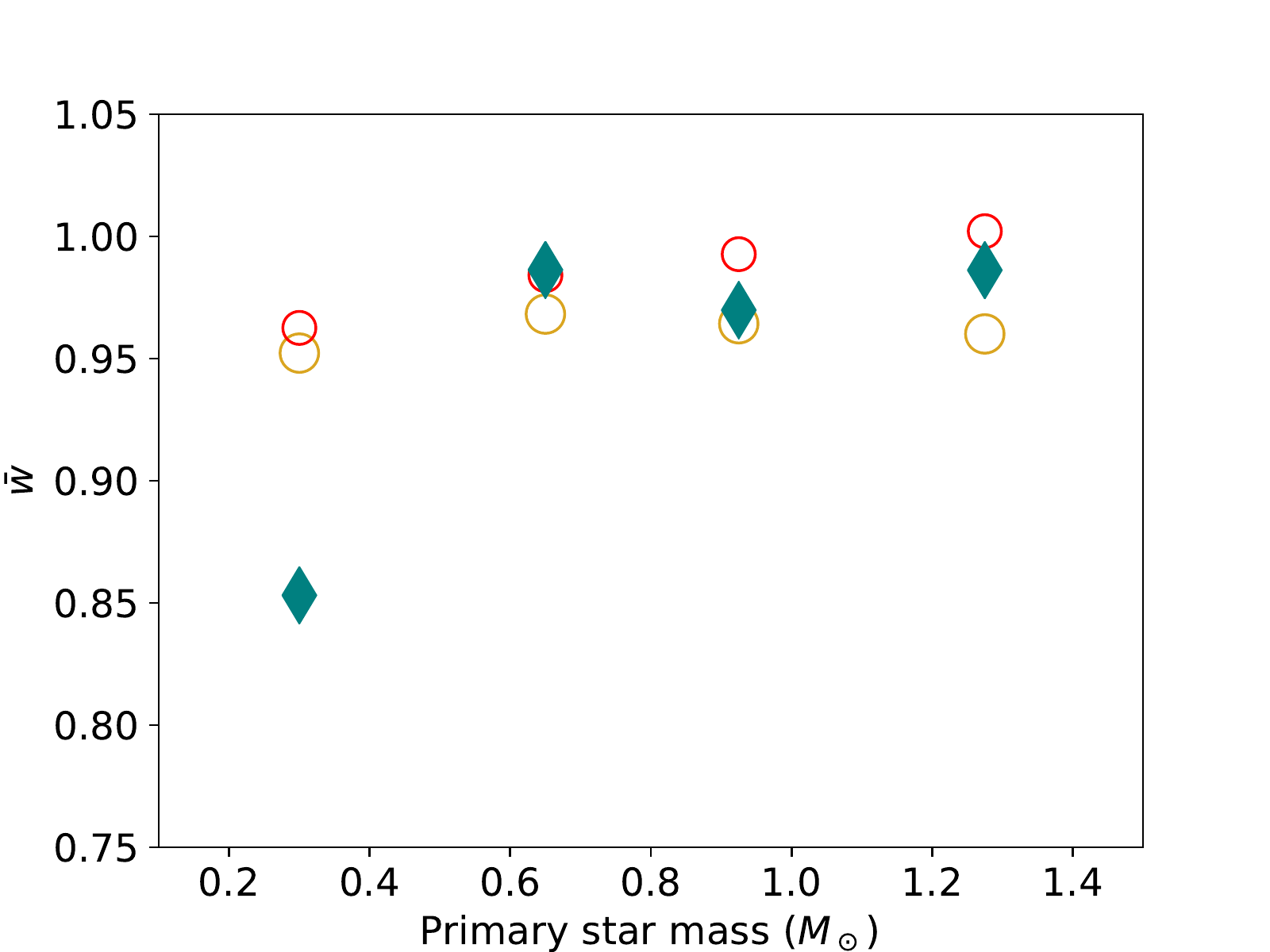}
\caption{Indices of habitability of circumstellar and circumbinary regions averaged in four bins of mass  corresponding to the range of masses of M-, K-, G- and F-type stars. Left panel: big and small circles: fraction $f_\text{hab}$ of habitable circumstellar regions around the primary and secondary star, respectively; diamonds: fraction $f_\text{P}$ of habitable circumbinary regions. Right panel: big and small circles: mean normalized widths $\overline{w}_\text{S}$ of the circumstellar HZs around the primary and the secondary, respectively; diamonds: mean, normalized width $\overline{w}_\text{P}$ of the circumbinary HZs. Results obtained from Model A (Table \ref{tab:results}). 
\label{fig_fhab_m}}
\end{figure}

By applying our methodology we can quantify the habitability of binary systems as a function of stellar mass. In Fig. \ref{fig_fhab_m}, we show the mean values of the indices of habitability $f_\text{hab}$ and $\overline{w}$ averaged in four bins of mass of the primary, $m_A$, corresponding to the range of masses of M-, K-, G- and F-type stars. In the left panel we show the fraction of habitable regions $f_\text{hab,SA}$, $f_\text{hab,SB}$ and $f_\text{hab,P}$ and in the right panel the normalized width of the HZ, $\overline{w}_\text{SA}$, $\overline{w}_\text{SB}$, and $\overline{w}_\text{P}$. The left panel shows that the number of circumstellar HZs decreases with increasing stellar mass, whereas the number of  circumbinary HZs increases with stellar mass. The right panel shows that at higher stellar masses there is an increase of the normalized width of circumbinary HZs. The strong dependence of stellar luminosity on stellar mass provides an interpretation of these trends in terms of an outwards shift of the radiative HZ (Fig.\,\ref{HZtypes}). Indeed, as the radiative HZ shift outwards, we expect a  larger chance of of type-3 conditions in circumstellar regions (e.g., a decrease of habitable fractions $f_\text{hab,SA}$, $f_\text{hab,SB}$) and a larger chance of type-1 habitability in circumbinary regions (i.e, an increase of $f_\text{hab,P}$). The probability of finding habitable planets in S-type orbits increases for late-type host stars. Instead, the chance of finding habitable planets in P-type orbits increases for early-type host stars. Binary systems with $m_A > 1.5$ M$_\sun$, not treated in this work, would probably have better circumbinary habitability conditions. 

\subsubsection{Trends with stellar metallicity}  
\label{sectTrendMetallicity}

Our results indicate that, for a given set of binary parameters, a decrease of stellar metallicity tends to give  a lower habitability of circumstellar regions and a higher habitability of circumbinary regions. This can be seen in Table \ref{tab:results} by comparing the results obtained from models with identical distribution functions, but with different stellar metallicity. One can see that a decrease of metallicity from $Z=0.017$ (the solar value) to $Z=0.008$ yields lower fractions of habitable circumstellar regions $f_\text{hab,SA}$ and $f_\text{hab,SB}$ and slightly higher fractions of habitable circumbinary regions $f_\text{hab,P}$. These results can be interpreted considering that, for a given stellar mass, the luminosity in the main sequence increases with decreasing metallicity \citep[see, e.g.,][]{Bressan12}. As a result, the radiative HZ shifts outwards, yielding a  larger fraction of circumstellar conditions of type 3, as well as a larger fraction of circumbinary conditions of type 1. In all cases, the differences that we find are very small, but systematic. Concerning the average normalized widths of the HZs, we find slightly larger values of circumbinary widths $\overline{w}_\text{P}$ at low metallicity, and negligible differences in the circumstellar widths $\overline{w}_\text{SA}$ and $\overline{w}_\text{SB}$.

In this context it is interesting to compare the results obtained from models with IMF and low-metallicity, representative of binary systems of an older stellar population (e.g.~models E and G), and models with PDMF and solar metallicity, representative of present-day stellar populations (e.g.~models B and D). All other model assumptions being the same, the results shown in Table \ref{tab:results} show that in the older stellar population (e.g.~model E) the circumstellar habitability decreases and the circumbinary habitability increases compared to the present-day population (e.g.~model B). The combined effect of changes of the mass distribution function, discussed in Section \ref{sectDistributionFunctions}, and metallicity tend to reinforce these systematic trends.

\begin{deluxetable}{cccccccc}[b!]
\tablecaption{Impact of planetary orbital eccentricity, $e_p$, on the indices of habitability $f_\text{hab}$ and $\overline{w}$ calculated with Model A.}
\tablecolumns{8}
\tablehead{ \colhead{Model$^a$} & \colhead{$e_p$} & \colhead{$f_\text{hab,SA}$} & \colhead{$\overline{w}_\text{SA}$} & \colhead{$f_\text{hab,SB}$} & \colhead{$\overline{w}_\text{SB}$} & \colhead{$f_\text{hab,P}$} & \colhead{$\overline{w}_\text{P}$} }
\startdata
A & 0.0 & 0.843 & 0.955 & 0.869 & 0.970 & 0.035 & 0.930 \\
A & 0.3 & 0.841 & 0.954 & 0.868 & 0.969 & 0.036 & 0.936 \\
A & 0.6 & 0.832 & 0.951 & 0.860 & 0.966 & 0.038 & 0.942 \\
\enddata 
\tablenotetext{a}{Model A is defined in Table \ref{tab:results}. Results obtained for $N_\circ = 10^5$.}
\label{planetaryEccentricity}
\end{deluxetable}

\begin{figure}
\includegraphics[width=1.0\columnwidth]{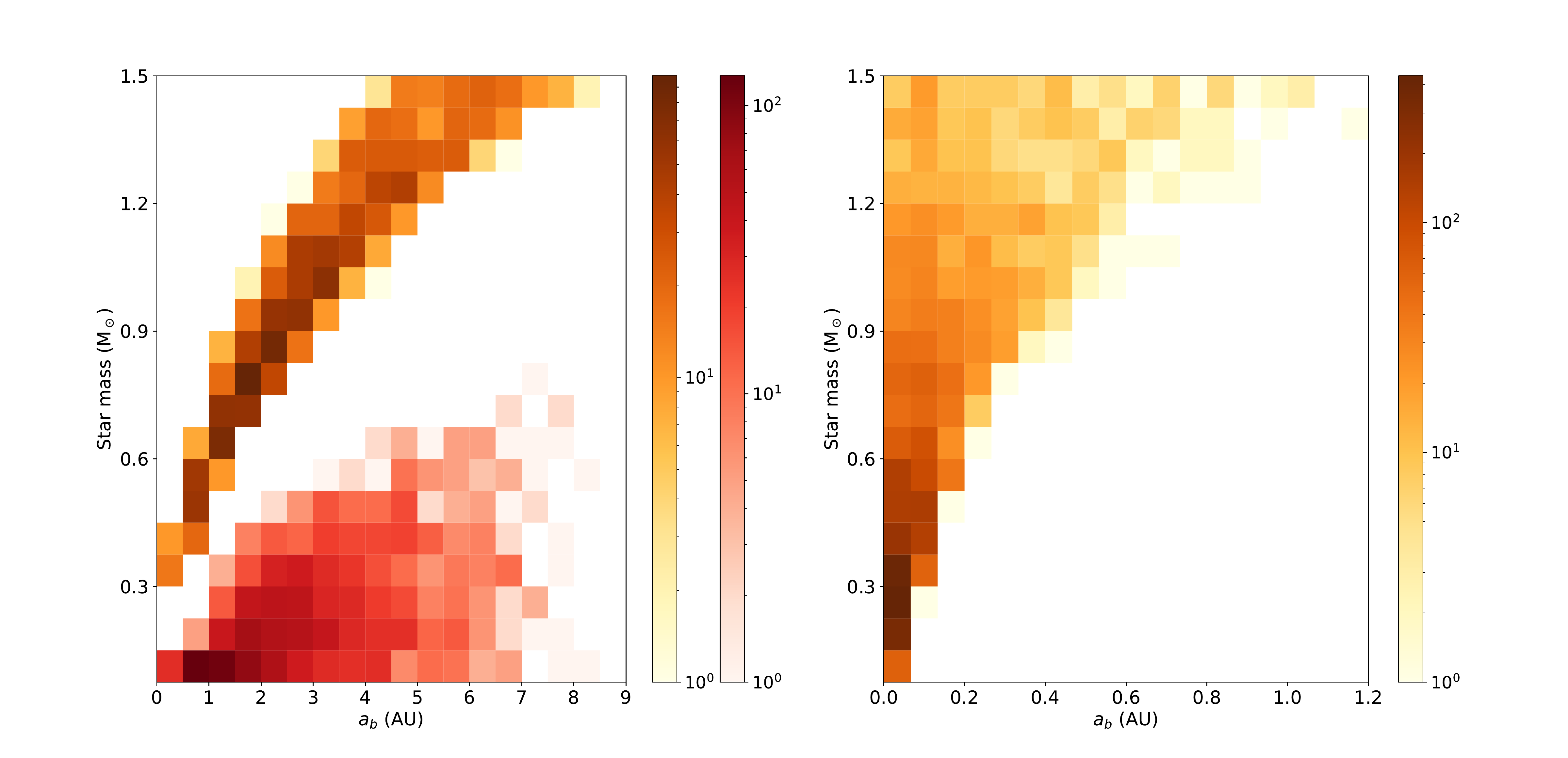}
\caption{Density plot of star masses versus binary semimajor axis for selected samples of binary systems with optimal conditions of circumstellar (left) and circumbinary (right) habitability. Left panel: systems with $w_\text{SB}>1.3$. Right panel: systems with $w_\text{P}>1.3$. Yellow bins refer to the primary star mass, red bins refer to the secondary star mass. Sample obtained from a simulation with $N_\circ=10^6$ of Model A (Table \ref{tab:results}). 
\label{fig_obs_limits}}
\end{figure}
 
\subsection{Searching for habitable planets in P-type orbits}
\label{sectSearchingHabExoplanets}

\subsubsection{Optimal conditions for circumbinary habitability}

The statistical results displayed in Table \ref{tab:results} predict a very small fraction of binary systems with circumbinary HZs. The main reason for this is the extremely wide distribution of orbital separations exhibited by binary stars, which can attain values up to $\sim 2 \times 10^5$ AU \citep[e.g.][]{Hartman20}. However, if we focus on close binaries ($\log a_b < -1$), most ($\sim 75\%$) of the binary systems allow the existence of type-1 circumbinary regions (Fig.~\ref{fig_norm_hist_ab}, right panel). Type-2 regions are possible for binaries with a separation of $\log a_b < 0.5$ ($P \lesssim 1000$ days for FGK-type twins), corresponding to 20\% of the separation distribution. Only a small fraction of the systems in the $-1< \log a_b <0.5$ range has a combination of stellar and orbital parameters that actually supports habitability. As for the eccentricity, Fig.~\ref{fig_norm_hist_eb} clearly indicates that even a small increase above 0 strongly reduces the fraction of systems hosting circumbinary HZs. Therefore, we can identify two distinct populations: one of close, circularized binaries ($P<30$ days), supportive of habitable circumbinary planets, and the other of wide, eccentric binaries ($P\ge 30$ days), mostly uninhabitable. On top of that, studies of planetary formation in circumbinary environments have raised concerns on the effective ability to form planets for binaries with $P \gtrsim 50$ days (see Section \ref{sectPlFormation_P}).

We can also observe (Fig. \ref{fig_wSB}, right panel) that, for a small subset of the type-1 and type-2 binaries, the supplemental insolation of the secondary can produce a habitable zone wider than that allowed around a single star with the same mass of the primary star. The maximum enhancement of the width of the habitable zone (+41\%) occurs for "twin-star" systems, where both stars have the same mass and hence the same luminosity. These twin-star, very close binaries have been already identified as especially conductive to life \citep{Shevchenko17} and our results support this conclusion, even though we must note that these systems seem to be very rare.

Concerning the mass of the primary star, Fig.~\ref{fig_fhab_m} casts light on the requirements for optimal circumbinary habitability. While extended habitable zones are not a prerogative of massive primaries, it is clear that the strict binary separation prerequisite relaxes as mass increases. Higher mass primaries are able to sustain an appropriate level of insolation at larger distances, which in turn allow for less closed secondaries. In fact, all type-1 systems with $\log a_b >-0.5$ require a primary with $\text{M}>0.8 \,\text{M}_\sun$, namely a G-type star. The extension of the circumbinary HZ is favoured when we consider higher mass primaries. This can be seen in Fig.~\ref{fig_obs_limits} (right panel), where we have selected the systems for which the circumbinary HZ is at least 1.3 times wider than in the case of a single central star with mass of the primary. Therefore, the mass of the primary can be considered a relevant statistical booster in the context of a rare scenario.

To summarize, even if the absolute number of binary systems with circumbinary HZs is low,  by selecting close ($P<10$ days), equal-mass binaries with G- and F-type primaries and low binary eccentricity we will find all the ingredients for increasing the circumbinary habitability.

\subsubsection{Lessons from the current sample of exoplanets in P-type orbits}

\label{sect_Pbiases}

So far, there are 33 known binary systems with substellar companions in circumbinary orbits (The Extrasolar Planet Encyclopedia\footnote[6]{http://www.exoplanet.eu}): 12 of them have been detected via timing methods (applied to eclipsing binaries and pulsars), 10 via transit photometry, 8 via direct imaging, 2 via to microlensing and 1 thanks to radial velocity. When we exclude from the computation the objects with mass -or minimum mass- over 13 $M_J$ (brown dwarfs), the detection by transit shrinks to 7, those by imaging to 4 and that by radial velocity (RV) to 0. In the case of Doppler surveys, past and present efforts have achieved typical measurement precision $> 20$ m s$^{-1}$ (e.g., Martin et al. 2019, and references therein). The circumbinary planets uncovered by the transit method have typical masses akin to that of Saturn. The amplitude of the RV reflex motion induced by such companions at orbital distances beyond the critical value prescribed by dynamical stability considerations straddles the threshold of detectability for RV times-series with the above mentioned precision level. This might explain the lack of RV detections for this sample.
Only 10 out of 33 systems are composed by Main Sequence stars and have their stellar and orbital parameters sufficiently constrained to conduct an analysis of their habitability. The circumbinary planets found around these systems have all been detected via the transit method. We now discuss the properties of this specific sample.

As can be seen in Table \ref{tab:keplers}, the ten systems are well characterized and somewhat homogeneous in their properties. They all have separations comprised between 0.08 and 0.23 AU. Apart from Kepler-34, which exhibits a binary eccentricity of 0.52, they all have eccentricities lower than 0.16. Eight out of ten systems have a G-type primary (Kepler-16 has a K-type primary and Kepler-1647 has a F-type primary). Finally, their mass ratios are relatively high, with the largest disparity sported by Kepler-453 with $q=0.21$.  In other words, they generally comply with the conditions for finding circumbinary HZs (low $a_b$, low $e_b$, and relatively early-type primaries) and for widening the circumbinary HZs (relatively high mass ratios). The detailed analysis of their circumbinary habitability confirms the existence of an overlap between radiative HZs and the regions of dynamical stability (Fig. \ref{fig_kepl_sys}).

At first sight, the fact that all the systems in Table \ref{tab:keplers} host circumbinary HZs seems at odds with the prediction that circumbinary HZs are extremely rare (Table \ref{tab:results}). Even when we consider the limits on $a_b$ for planetary formation and reduce the fraction of binaries with red dwarf primaries, we find that at least $1/2$ of the systems does not support type-1 or 2 circumbinary HZs (Table \ref{results_sub}).
The detections collected in Table \ref{tab:keplers} point to the existence of observational biases that favor the detection of circumbinary planets in systems where the circumbinary habitability is possible.
One obvious bias is that exoplanets are expected to be found in stable orbits, which means that circumbinary exoplanets must lie beyond the outer boundary of dynamical stability. 
At the same time, the geometrical probability of detecting exoplanets with the transit method increases with decreasing semimajor axis of the planetary orbit. The combination of these two biases means that circumbinary exoplanets can be detected with the transit method only if the outer boundary of dynamical stability, $a_\text{crit,P}$, is sufficiently close to the stars. In turn, according to Eq. \eqref{acritP}, the vicinity to the stars of $a_\text{crit,P}$ implies that $a_b$ must be small, which is exactly the condition that maximizes the existence of circumbinary HZs.

A more subtle bias emerges from the study of circumbinary planets around eclipsing binaries. It has been noted that this subset of planet-hosting binary systems is characterized by statistically larger than average periods \citep{Martin15,Martin18}. A proposed explanation of this bias involves the presence of a third, undetected faint companion in a large fraction of these systems. Another hypothesis by \citet{Fleming18} explains it through an initial angular momentum transfer that widens a tight binary during the pre-MS stage and a successive magnetic braking that shrinks again the system. During the first stages of this process the closest circumbinary planets  would be ejected. As a result, circumbinary planets would only be  detected in wider systems ($P_b \gtrsim 5$ days), where this complex physical mechanism does not take place. If the systems have $P_b \lesssim 30$ days our predictions show that circumbinary HZs can still exist and be populated. 

Finally, we note that the sample of known circumbinary planets is  biased against the detection of small ($<3\,M_\earth$) objects. Detecting these planets via the transit method presents observational challenges, as the very shallow events exhibit transit timing variations (TTVs) on the order of days in magnitude, irregular depths and durations \citep{Armstrong13,Martin20}.

\subsubsection{Analysis of the known circumbinary planet-hosting binaries}

We now discuss the known circumbinary planet-hosting systems selected in the previous section one by one. Many of them have the habitability of their known circumbinary planets already explored \citep{Haghighipour13}. \citet{Quarles18} have assessed the possible existence of other, more interior planets in these systems, while \citet{Barbosa20} discussed the formation of other planets besides those already discovered simulating their protoplanetary disk stage. Here, we apply Eq.~\eqref{DELTAhabP} to insert them into the framework of our analysis of habitability.
In addition, we investigate the possibility that undetected planets with terrestrial size may exist in the circumbinary habitable region.

The possible planet-planet interactions, that must be considered in order to evaluate the potential presence of Earth-sized planets in these systems, are treated under the conclusions derived by \citet{Quarles18}. For each known circumbinary exoplanet, they defined a range of semimajor axes that would give rise to repeated close planetary encounters and the eventual destruction of the system architecture. These values are tested in massive numerical simulations and expressed in terms of planet-planet mutual Hill spheres \citep{Chambers96}:
\begin{equation}
R_{H,m}=\frac{a_1+a_2}{2}\biggl(\frac{m_1+m_2}{3M_{int}}\biggl)^{1/3}
\end{equation}
where $a_1$, $a_2$, $m_1$ and $m_2$ are, respectively, the semimajor axes and the masses of the innermost and outermost planet considered and $\text{M}_{int}$ is the total mass interior to the outermost planet. The authors found that, in order to attain the long-term stability in all their tests, $a_2-a_1>\beta \times R_{H,m}$ with $\beta=7$. This is a very conservative estimate, as they noted, and in some circumstances $\beta$ can be as low as 5. However, in this work we will use the former value. This method gives us the opportunity to make a first-order assessment of the ability of a given binary system to host another planet in its HZ.

\begin{figure}
\includegraphics{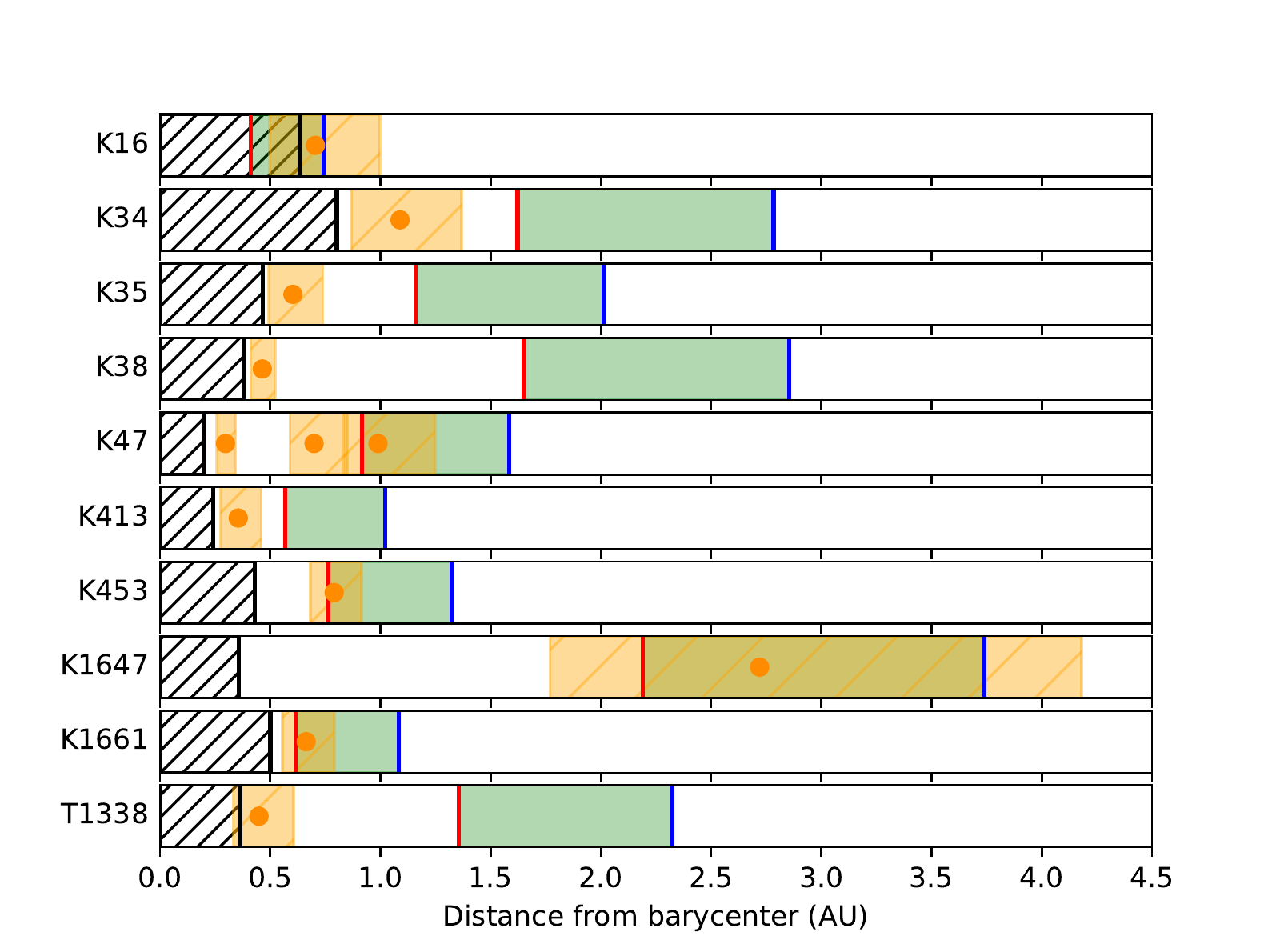}
\caption{Combined analysis of the circumbinary regions hosting the Kepler and TESS exoplanets discovered so far. Orange dots: detected planets. Black hatched areas: region of dynamical instability. Red and blue lines: inner and outer edges of the radiation HZs.  Orange areas: ranges of semimajor axes that cannot continuously host an additional Earth-sized (or larger) planet under the conditions derived by \citet{Quarles18}.}
\label{fig_kepl_sys}
\end{figure}

The combined results of our analysis of dynamical stability and habitability are summarized in Fig.\,\ref{fig_kepl_sys}. We now discuss individual cases. 

\begin{deluxetable}{lccccl}[b!]
\tablecaption{The 10 currently known Main Sequence binary systems that host circumbinary planets. The masses in the second and third column are in solar units and the semimajor axis in the fourth column is in AU. It should be noted that other works include also Kepler-64 but, being actually a 4-component system, is not discussed in our study.\label{tab:keplers}}
\tablecolumns{6}
\tablehead{\colhead{Name} & \colhead{$M_{A}$} & \colhead{$M_{B}$} & \colhead{$a_b$} & \colhead{$e_b$} & \colhead{ref}}
\startdata
Kepler-16   &  0.689 & 0.203 & 0.224 & 0.159 & \citet{Doyle11} \\
Kepler-34   &  1.048 & 1.021 & 0.229 & 0.521 & \citet{Welsh12} \\
Kepler-35   &  0.890 & 0.810 & 0.176 & 0.142 & \citet{Welsh12} \\
Kepler-38   &  0.949 & 0.249 & 0.147 & 0.103 & \citet{Orosz12b} \\
Kepler-47   &  1.043 & 0.362 & 0.084 & 0.023 & \citet{Orosz12a} \\
Kepler-413  &  0.820 & 0.542 & 0.102 & 0.037 & \citet{Kostov14a,Kostov14b} \\
Kepler-453  &  0.944 & 0.195 & 0.185 & 0.052 & \citet{Welsh15} \\
Kepler-1647 &  1.221 & 0.968 & 0.128 & 0.159 & \citet{Kostov16} \\
Kepler-1661 &  0.841 & 0.262 & 0.187 & 0.112 & \citet{Socia20} \\
TOI-1338    &  1.038 & 0.297 & 0.129 & 0.156 & \citet{Kostov20} \\
\enddata
\end{deluxetable}

Kepler-16 is the least habitable system in the sample, having a type-2 HZ with a reduced extension, $w_\text{P}=0.29$, compared to a single-star HZ. This is caused by the interplay between  the relatively large separation in this sample (0.22 AU) and the low luminosity of the two central stars. The system hosts a 0.33 $\text{M}_{jup}$ planet right in the middle of its HZ, that makes quite unlikely the presence of another, Earth-sized planet in it. 

Kepler-38, 47, 413, 453, 1661 and TOI-1338 have type-1 HZs, with $w_\text{P}$ in the range from 1.0 to 1.08. These systems should present an environment remarkably similar to that of our own Solar System, especially when we consider that they all have G-type primaries. Kepler-38, Kepler-413 and TOI-1338 host respectively a 5.1 $\text{M}_\earth$, a 0.21 $\text{M}_{jup}$ and a 0.32 $\text{M}_{jup}$ planet, all of them orbiting close to the inner edge of the respective dynamical stability zones and leaving the habitable zones entirely free for other possible bodies. Kepler-47 has three known planets with masses between 9 and 20 $\text{M}_\earth$, two of them out of the HZ, being too close to the central stars to be habitable, and the third in the HZ but near its inner edge, possibly leaving room for another, cooler body in an outer orbit. Kepler-453 and Kepler-1661 both present a similar situation, hosting respectively a $10$ $\text{M}_\earth$ and a $16$ $\text{M}_\earth$ planet in their HZ, near the inner edge.

Finally, Kepler-34, 35 and 1647 have extended type-1 HZs with $w_\text{P}$ in the range from 1.1 to 1.36. Kepler-34 and Kepler-35 host respectively a 0.22 and a 0.13 $\text{M}_{jup}$ planet, both of them orbiting between the dynamical stability edge and the inner edge of the HZ. Once again, their HZ seem able to host Earth-sized planets. Considerations about disruption by planetary migration are not discussed here, but we note that such a mechanism may have prevented the formation or eliminated possible habitable bodies. Kepler-1647 hosts a 1.52 $\text{M}_{jup}$ planet in the central region of the HZ.

\subsection{Searching for habitable planets in S-type orbits}

\subsubsection{Optimal conditions for circumstellar habitability}
\label{sect.optimalS}

In contrast with the results related to circumbinary habitability, our Monte Carlo sampling (Tables \ref{tab:results} and \ref{results_sub}) shows that planets in circumstellar orbits should encounter favourable conditions in most cases. The explanation for this fact is once again related to the extremely wide distribution of binary semimajor axis. As it is possible to see in Fig. \ref{fig_obs_limits}, the radiative influence of the secondary star rapidly decreases with increasing binary separation and for most of the systems is expected to be negligible. In particular, if we discriminate the systems where the circumstellar habitability conditions substantially deviates from the single-star case when $w > 1.1$, we find that only $\sim 1.2\%$ of the type-1 and 2 HZs around secondary and no HZ around primary fall in this category. If we raise the threshold to $w > 1.3$ (as in plot \ref{fig_obs_limits}), the percentage of HZs around secondary reduces to $\lesssim 0.3\%$. In other words, we expect that most of the circumstellar planets in binary systems are subject to the same conditions of those in orbit around single stars.

On the other hand, close binary systems offer a range of exotic conditions that could result in vast enhancements of habitability, especially around secondary stars (as discussed in \ref{subsectSecondaryHab}). This subset of binaries is also responsible for most of the trends that we observe in our habitability indexes. As it is possible to see in Fig.\,\ref{fig_fhab_m} (left panel), circumstellar habitability is favoured around low mass stars, both for the primary and the secondary. In particular, the mean extension $\overline{w}$ around the secondary is the largest on the entire mass range (right panel), although it must be noted that $\overline{w}$ is high in all cases, the lowest  value being $\simeq 0.9$. Fig.\,\ref{fig_norm_hist_mu} (left panel) hints that type-1 habitability around primary stars increases with the mass ratio between the components, while type-1 habitability around the secondary is substantially flat. A clearer trend is visible in Fig.~\ref{fig_norm_hist_eb}, where type-1 habitability around both the primary and the secondary component is favoured by low eccentricies. At exactly $e=0.0$ the fraction of type-1 HZ decreases significantly, but this is the effect of the correlation between eccentricity and orbital period introduced by the existence of a circularization period (see sect.~\ref{sectPeriods}): a consistent fraction of zero-eccentricity systems are also very close binaries that do not present dynamically stable S-type orbits sufficiently far away to the star. If only systems with $P_\text{b}>P_\text{circ}$ are included, the increasing trend in type-1 HZ width with decreasing eccentricity is uninterrupted. To conclude, apart from the wide binaries whose circumprimary or circumsecondary orbit can be studied as separated single stars, the best places for searching habitable exoplanets seems to be low mass systems with moderate separations and low eccentricites, that can offer enlarged HZs around secondary stars.

Even if the circumstellar habitability inside binary systems might not be impeded by dynamical instability, there are concerns regarding the disruption of the protoplanetary disk during the formation phase, as discussed in Section \ref{sectPlFormation_S}.

\subsubsection{Analysis of the known circumstellar planet-hosting binaries}

Currently, there are 90 known binary systems with projected separation under 1000 AU that host at least one planet in S-type orbit (the Extrasolar Planet Encyclopedia and the \citet{Schwarz16} catalogue). Among them, 48 have been detected via radial velocity, 39 via primary transit photometry, 2 via microlensing and 1 via astrometry. This sample is larger than that of known circumbinary planets, but the stellar and orbital properties of these systems are somewhat less constrained. Most (63) of these systems have projected separations above 100 AU, possibly indicating a bias in favor of wide binaries. A physical explanation of this bias points to the difficulty to form planets in the gravitationally active environment influenced by the presence of a close companion star \citep{Kraus16}. An explanation in terms of observational bias is that, for such large separations, the reflex motion induced by one star on the other is small and varies very little in time. This makes the probability of detecting exoplanets around any of the components no different from the probability of detecting exoplanets around single stars. In fact, in some instances, the binarity has been discovered after the detection of the planet. This fact makes also clear that this statistics is far from definitive. On one side, there are ongoing searches \citep[e.g.][]{Ziegler17} for possible companions of stars currently considered single. On the other, some of the systems currently considered binary might be part of higher order multiplicity groups.

As explained in sect.~\ref{sect.optimalS}, the influence of the one star on the other's HZ rapidly decreases with increasing separations. Therefore, in order to give them a closer look in the framework of our paper, we have selected the 12 known binaries with $a_b<30$ AU. We then excluded 5 of them for the lack of sufficient data (OGLE2013-BLG0341 and HD 7449), the presence of a non-Main Sequence primary (HD 59686 and HD 87646) and the misalignment of planetary and companion star orbital planes (Kepler-693). The 7 remaining systems and the method used for the detection of their planets are reported in table \ref{tab:stype}.

All the selected systems have late-type (M- or K- type) secondaries and relatively large  eccentricities. Even if these conditions tend to favor circumstellar habitability, the effects of binarity on these systems are generally negligible. An enhancement in the order of $1-3\%$ is present in the widths of the secondary star HZs in 3 systems, while for the primary star HZs the effect is $<0.2\%$ in every instance. Only the two systems with small stellar separations (HD 42936 and Kepler-420) show significant effects of binarity on their HZs. This fact stresses, again, the dominant role of stellar separation on the habitability  of binary systems.  
In Kepler-420 the gravitational pull of the secondary reduces $w_\text{SA}$ to 0.16, making it an uncommon example of type-2 HZ. Instead, the secondary star HZ is slightly enhanced. Finally, in HD 42936 the secondary orbits in the outer part of the primary star HZ. Therefore, there is no room for habitable planets around the most massive component. On the other hand, the secondary star has an enhanced type-2 HZ. The flux from the primary generally dominates the energy balance in the secondary HZ, and accounts for $\sim 47\%$ of the total energy that would be received by an hypothetical planet at exactly the inner edge of the HZ.

\begin{deluxetable}{lcccccccl}[b!]
\tablecaption{The 7 binary systems with circumstellar exoplanets currently known with component separations $< 30$ AU, Main Sequence components and sufficiently constrained stellar and orbital parameters. Masses are in solar units, $a_b$ is AU.\label{tab:stype}}
\tablecolumns{9}
\tablehead{\colhead{Name} & \colhead{$M_{A}$} & \colhead{$M_{B}$} & \colhead{$a_b$} & \colhead{$e_b$} & \colhead{$w_\text{SA}$} & \colhead{$w_\text{SB}$} & \colhead{Method$^b$} & \colhead{ref}}
\startdata
HD 41004      &  0.7  & 0.4  & 23  &$0.2^a$& 1.00 & 1.00 & RV & \citet{Zucker04} \\
HD 196885     &  1.33 & 0.45 & 21   & 0.42 & 1.00 & 1.02 & RV & \citet{Correia08} \\
$\gamma$ Cephei &  1.41 & 0.41 & 20.2 & 0.41 & 1.00 & 1.03 & RV & \citet{Hatzes03} \\
Gliese 86     &  0.83 & 0.49 & 19   & 0.40 & 1.00 & 1.00 & RV & \citet{Queloz00} \\
HD 176051$^c$     &  1.07 & 0.71 & 18.9 & 0.54 & 1.00 & 1.01 & AS & \citet{Muterspaugh10} \\
Kepler-420$^d$    &  0.99 & 0.70 & 5.3  & 0.31 & 0.16 & 1.01 & TR & \citet{Santerne14} \\
HD 42936      &  0.87 & 0.08 & 1.22 & 0.59 & 0.00 & 1.14 & RV & \citet{Barnes20} \\
\enddata
\tablenotetext{a}{Assumed eccentricity.}
\tablenotetext{b}{RV: radial velocity, AS: astrometry, TR: transit.}
\tablenotetext{c}{Unconfirmed.}
\tablenotetext{d}{The orbital properties of the binary companion, derived for an assumed distance of $\sim900$ pc \citep{Santerne14}, are uncertain. The Gaia DR2 astrometric solution reports a negative parallax, it has a large excess noise (2.44 mas), and the reduced unit weight error (RUWE) statistics is $\simeq12$, indicating an unreliable solution (well-behaved solutions being those with $\text{RUWE} < 1.4$ \citep[see e.g.][]{Lindegren18}. This might be due to a combination of unmodeled orbital motion and confusion due to light contribution from the binary companion.}
\end{deluxetable}

\subsubsection{Searching for habitable planets around the secondary}
\label{subsectSecondaryHab}

The enlargement of the HZ around the secondary star (left panel in Fig. \ref{fig_wSB}), resulting from the outwards shift of the outer edge of the radiative HZ (Section \ref{sec:results}), is potentially interesting for observational searches of habitable planets in binary systems. For this reason, we performed several tests to understand in which range of binary parameters this effect is prominent. The only significant trends we found are shown in the left panel of Fig. \ref{fig_obs_limits}, where we plot the systems for which the HZ is extended by more than $30\%$ than the HZ of a single star of the same mass. One can see that the enlargement of the HZ takes place for small values of binary semi-major axis ($a_b \lesssim 8$ AU),  when secondary stars are of M type (red dots with $m_B \lesssim 0.6$ $\text{M}_\sun$), and when the primaries have a well-defined range of masses as a function of $a_b$ (yellow dots). Even if the fraction of systems with these properties is small ($\sim 0.1-0.3$\% across different models), it would be extremely interesting to search for extended HZs around M-type stars. Since the enlargement of the HZ is due to the outward shift of the outer edge of insolation, with a proper selection of  binary systems this configuration may provide a unique possibility to find habitable planets not affected by tidal locking around M-type stars. 

\section{Conclusions}
\label{sec:conclusions}

We have designed a suite of Monte Carlo experiments aimed at generating large samples of binary stellar systems. The systems are extracted by random sampling of a set of PDFs of stellar masses and orbital parameters representative of binary stars in the main sequence, with masses in the range from 0.08 to 1.5 $\text{M}_\sun$. The simulated samples of binary stars are used to quantify the statistical properties of planetary habitability in binary systems. For each extracted system, we estimate the luminosity and effective temperature of the primary and secondary star using stellar evolutionary tracks calculated at two  values of stellar metallicity (solar and 1/2 solar). From the dynamical data of each system and the application of algorithms obtained from N-body simulations \citep{Quarles18,Quarles20} we calculate the location and extension of regions of dynamical stability for planetary orbits. By weighting the radiative flux received by the two stars \citep{Kaltenegger13,Haghighipour13} we calculate the boundaries of the radiative HZs. We define three possible conditions of long-term habitability from the study of the intersection between dynamically stable regions and radiative HZs in the stellar orbital plane. If the intersection is complete, partial or null, the habitability is classified of type 1, 2, and 3, respectively. This definition is applied both to circumstellar, S-type orbits and to circumbinary, P-type orbits. To gauge the impact of stellar binarity on habitability, the extension of type-1 and type-2 HZs is compared to the extension of the classic HZ around a single star with the same mass of the host star. To assess the robustness of the results, the Monte Carlo experiments are repeated for different models, each model being defined by a specific set of PDFs. The results obtained from different models consistently support the following conclusions.

\begin{itemize}

\item
In $\sim 95-97$\% of all binary systems, circumbinary habitability is prevented by the lack of overlap between the radiative HZ and the dynamically stable region (type-3 condition). However, significant fractions of circumbinary HZs exist at the low-end tail of the distribution of stellar separations, when $a_b \lesssim 0.3$\,AU. 

\item
Circumstellar habitability is possible for $\sim 78-94$\% of binary systems, in most cases with a complete overlap between the radiative HZ and the dynamically stable region (type-1 condition). However, habitable circumstellar regions become rare when $a_b \lesssim 1$\,AU.


\item

If we consider the sub-samples of systems where planetary formation is currently believed to be possible, we find that $99-100\%$ of systems with $a_b>10$\,AU can feature circumstellar HZs, whereas $16-23\%$ (or $35-44\%$) of systems with $a_b<0.3$\,AU (or $<1.0$\,AU) can feature circumbinary HZs. 
If we consider also a lower limit $P_b > 5$ days for circumbinary planetary formation, we find a reduction of $20-25\%$ of the circumbinary habitable fraction.
Therefore, the conditions for circumbinary habitability are stringent even when planetary formation is taken into account. 


\item
The habitability fractions are marginally affected by different hypotheses on the occurrence rate of binaries with primary stars of different spectral types. A factor-of-two reduction of the systems with M-type primaries does not change the circumstellar habitability and increases the circumbinary habitability by $4-6\%$.

\item
Cases in which the radiative HZ is interrupted by the boundary of dynamical instability (i.e., type-2 conditions of habitability) are rare and are found around $a_b \sim 1$\,AU in circumstellar regions and around $a_b \sim 0.1$\,AU in circumbinary regions. 

\item
The eccentricity of stellar orbits, $e_b$, affects the habitability around binary stars. Both circumbinary and circumstellar habitability are favored at low $e_b$, but in the latter case this result is partially masked by the correlation between period and eccentricity introduced by the circularization period $P_\text{circ}$. This causes a drop at exactly $e_b=0.0$.

\item
Stellar hosts of F and G spectral types yield slightly higher fractions of habitable circumbinary regions than hosts of K and M spectral types. Instead, the fraction of habitable circumstellar regions tends to increase for host stars of later spectral type. 

\item
The impact of stellar metallicity on binary habitability is almost negligible, with a slight increase ($\lesssim 1$\%) of the fraction of habitable circumbinary regions when the metallicity decreases from solar to half the solar value. 

\item 
Owing to the radiative  contribution of the companion star, circumstellar HZs around M-type secondary stars can be several times (up to one order of magnitude) more extended than in the case of a single star. Circumbinary HZs can be up to 41\% more extended when the mass ratio of the two stars is $q=m_\text{B}/m_\text{A} \sim 1$.  

\item
The ranges of $a_b$ and stellar masses that allow the existence of extended HZs are well constrained (Fig. \ref{fig_obs_limits}) and can be used to focus searches for habitable planets in binary systems. An intriguing, albeit hard to detect, possibility would be to search for extended HZs around M-type stars, with the outer edge located beyond the tidal-lock limit. 

\end{itemize}

To set the predictions of our statistical analysis in an observational framework, we have investigated the properties of the binary systems where exoplanets have been discovered so far. Circumbinary HZs are found in all the systems that host transiting planets in P-type orbits. Different observational biases conspire to boost the detection of such systems despite the rarity of circumbinary HZs in the general population of binary stars.
In most of the cases, circumbinary HZs in these systems may potentially host additional Earth-sized planets not detected so far. As far as the exoplanets in S-type orbits, at present time the quality of the observational data of the binary systems hosting these exoplanets is insufficient to perform a general study of the habitability of circumstellar regions. 

Future acquisition of experimental data and progress in climate modeling will provide tighter constraints on the results obtained in the present work. For instance, the experimental determination of the PDFs of binary stellar masses and orbital parameters is expected to improve significantly based on results from ESA's Gaia mission. The next Gaia intermediate data release (DR3), planned for the first half of 2021, will contain for the first time information on large numbers of stellar binaries detected in astrometry, spectroscopy, and photometry and this, combined with updated information on resolved physical binaries, will allow to access with improved statistics a very wide range of orbital separations and mass ratios \citep[e.g.][]{Soderhjelm04,Soderhjelm05}. As a result, it will be possible to obtain more accurate predictions from the Monte Carlo experiments that we have designed. With the use of dedicated climate models, it will be possible to better estimate the radiative boundaries of binary habitability by tracking the insolation received from both stars in the course of the planetary orbital motion. Preliminary work of this type, focused on some specific cases of circumbinary habitability \citep{Forgan14,Forgan16,Popp17}, should be extended to cover a broader palette of conditions representative of S-type and P-type orbits. 

Future observations of exoplanets in binary stars with ground- and space-based facilities (e.g., extreme-precision spectrographs such as ESPRESSO, planetary transit missions such as PLATO) may lead to the discovery of terrestrial-type planets in particular configurations of habitability that are predicted to exist from our study. Considering the high frequency of stellar binaries, the comparison between large observational data sets and the statistical predictions obtained with our methodology will provide a valuable tool to study the potential distribution of habitable planets and life in the Galaxy.

\acknowledgments

The authors thank the anonymous referee for the very careful reading of the draft and her/his many useful suggestions.

P.S. wish to thank the European Space Agency for co-funding his doctoral project (EXPRO RFP IPL-PSS/JD/190.2016).
L.S. and G.V. wish to thank the Italian Space Agency for co-funding the Life in Space project (ASI N. 2019-3-U.0).
This work has made use of data from the European Space Agency (ESA) mission {\it Gaia} (\url{https://www.cosmos.esa.int/gaia}), processed by the {\it Gaia} Data Processing and Analysis Consortium (DPAC, \url{https://www.cosmos.esa.int/web/gaia/dpac/consortium}). Funding for the DPAC has been provided by national institutions, in particular the institutions participating in the {\it Gaia} Multilateral Agreement.

%





%

\end{document}